# Spatio-temporal Analysis for Extreme Temperature Indices over Levant region

Ala A. M. Salameh[1], Sonia Raquel Gámiz-Fortis[1], Yolanda Castro-Díez[1], Ahmad Abu Hammad[2] and María Jesús Esteban-Parra[1]

[1]*Departamento de Física Aplicada, Universidad de Granada.*

[2] *Department of Geography, Birzeit University*

alasalman84@correo.ugr.es, srgamiz@ugr.es, ycastro@ugr.es, ahammad@birzeit.edu, esteban@ugr.es

Short title: Analysis for Extreme Temperature Indices over Levant region

**Abstract**

The temporal and spatial trends of 16 climate extreme indices based on daily maximum and minimum temperatures during the period 1987–2016 at 28 stations distributed across the Levant region were annually and seasonally analyzed. The non-parametric Man-Kendall test and the Sen's slope estimator were employed for the trend analysis. Results showed that the region has significantly experienced a dominant warming trend for the last three decades, with more intense changes for minimum temperatures than for maximum. At annual scale, maximum values of minimum temperatures (TNx) exhibited significant increasing trends up to 0.68$^{0}$C/decade. For percentile-based extreme temperature indices, changes detected were more pronounced than those for the absolute extreme temperature indices, with 93% and 89% of stations significantly showed increasing trends in TX90p and TN90p, respectively. The duration and fixed threshold extreme indices confirmed the trend toward a warming, with the 86% of the stations exhibited significant increasing trends in the annual occurrence of summer days (SU25) and tropical nights (TR20). Moreover, 57% of stations showed significant increasing trends in their very summer days (SU30) index. At seasonal scale, the analysis of trends for extreme temperature indices showed intense and broad significant increasing trends in all absolute extreme temperature indices. In summer, more than 75% of total stations exhibited significant increasing trends for warm days and warm nights (TX90p and TN90p). In winter and spring, 71% of the total stations also showed significant increasing trends in SU25 index, whereas the percentage of stations reached 82% in summer and 64% in autumn for significant increasing trends in TR20 index. Finally, the influence of large-scale circulation patterns on temperature extremes was examined. The results highlighted the presence of significant correlations between most of the selected extreme temperature indices and the North Sea-Caspian pattern (NCP) at annual and seasonal scales.

## 1. Introduction

Climatic extreme events can have serious impacts on environment and society compared with the changes in the average climate (Vörösmarty et al., 2000; Parmesan et al., 2000). Furthermore, extremes are more sensitive to climate change than mean values (Katz and Brown, 1992). However, most analyses of long-term global climate change using observational temperature and precipitation data have focused on changes in mean values (Alexander et al., 2006), and there are few studies about climate extremes, especially in a region like Levant. The analysis of changes in extremes depends on daily data, which are generally difficult to access in countries like Jordan, Lebanon, Syria and Palestine, which impose restrictions on the accessibility of climatic data.

According to the studies of the Intergovernmental Panel on Climate Change (IPCC), the warming in southern and eastern Mediterranean over this century will be larger than the global average warming and the annual precipitation is very likely to decreases (IPCC, 2007, 2013). For example, Lelieveld et al. (2012) predict a continual, gradual and relatively strong warming of about 3.5-7°C between the period 2070-2099 and the 1961-1990 reference period and under an intermediate IPCC SRES scenario. Also, the Mediterranean experienced a downward precipitation trend through the second half of the twentieth century (Xoplaki et al., 2003), and this trend is expected to continue, with a decrease of up to 20% in total annual precipitation by the year 2050 (Black, 2009). The IPCC's scenarios for the eastern Mediterranean indicate that average summer temperatures will gradually increase around 0.5-0.9°C/decade, and the number of warm days will increase by 50-60 additional days/year by the end of the 21st century (Lelieveld et al., 2012). Note as is indicated by Lelieveld et al. (2012), the expected changes in precipitation could be associated with a reduction of cloudiness, which allows more solar radiation to be absorbed at the surface, being able to contribute to the increase of temperature. In addition, the Levant region in general, and Historical Palestine (Israel and Palestine), in particular, are considered more vulnerable and sensitive to the negative impacts of climate change, because these regions are affected by many developmental stresses such as the water resources shortage, weakness of existing infrastructure, low adaptive capacity, and frequent drought events. Moreover, all of these factors are escalated by the political conflicts and the rapid population growth (Hammad and Salameh, 2018; Al-Qinna et al., 2011; Ziv et al., 2006; Lelieveld et al., 2012; Sowers et al., 2011; Terink et al., 2013).

For Levant region, there are very few studies focused on changes in climate extremes indices. Zhang et al. (2005) analyzed changes in several indices over the Middle East region for the period (1964-1999), using data from 52 stations of 15 countries, but only 4 stations belong to Levant region. Donat et al. (2014) examined changes in extreme temperature and precipitation in the Arab region, but their study did not include stations from the Historical Palestine. However, changes in the frequency and intensity of climate extreme indices have been already well documented in other countries such as Saudi Arabia (Islam et al., 2015; Almazroui et al., 2014; Athar, 2014), Arabian Peninsula (AlSarmi and Washington, 2014),

Iran (Rahimi and Hejabi, 2018; Rahimzadeh et al., 2009) and Turkey (Toros, 2012; Erlat and Türkeş, 2013).

Therefore, the present study has a dual objective. Firstly, it tries to fill the gap in the studies related to extreme temperature indices in the Levant region. Secondly, it provides a comprehensive temporal and spatial analysis of extreme temperature indices and their relationships with the main large-scale circulation pattern in the Northern Atlantic and the Mediterranean Basin, i.e., the North Atlantic Oscillation (NAO), the East Atlantic (EA) pattern and the EA/Western Russia (EA/WR) pattern, the Western Mediterranean Oscillation (WEMO), the Mediterranean Oscillation (MO), and the North Sea-Caspian pattern (NCP).The study also includes an analysis of the El Niño-Southern Oscillation (ENSO) influence over Levant extreme temperature indices, as the main global climate variability pattern.

This study is organized as follow: Section 2 describes the study area, datasets, quality control and homogeneity tests utilized, as well as the methodology used. Results are presented in Section 3, while Section 4 shows a summary and discussion of the results.

## 2. Data and methods

### 2.1 Study area and temperature data

The geographical domain of this study extends over the entire Historical Palestine, that nowadays is composed by Israel and Palestine territories, located at latitude 29-33$^0$N and longitude 34-35$^0$E, with an area of about 27000 km$^2$ and elevation ranges between 392 m below the sea level and 1208 m above the sea level (Figure 1). The area lies in a transitional zone between the hot and arid southern part of west Asia and the relatively cooler and wet Mediterranean region (Maldonado, 2011).

Initially, the historical archive of daily time series of maximum and minimum temperatures from 118 stations was taken from the Israeli Meteorological Department (https://ims.data.gov.il). However, not all of these time series were used in the present study. They were subjected to an exhaustive data quality control (QC) to identify systematic errors, missing data and outliers (Klein Tank et al., 2009; Alexander et al., 2006; El Kenawy et al., 2013). In the first stage, time series were filtered based on the temporal continuity (at least 27 years) and coverage for the recent years. The stations with shorter records were not used for the computation of the extreme temperature indices but they were still useful for assessing missing data and outliers at nearby stations.

In a second stage, systematic errors, missing data and outliers were checked. Stations with more than 20% of missing data or incomplete years were discarded from data set (Klein Tank and Können, 2003). Only those months and years with more than 3 and 15 missing days, respectively, were handled. Note that the RClimDex v1.0 software used does not calculate monthly indices if more than 3 days are missing in a month, and if there are more than 15 missing days for the annual values. If these missing days were not successive the temporal interpolation was used based on the average two values of neighboring days (before and after the missing days). Furthermore, if these missing days were successive, the spatial interpolation method was applied based on neighboring stations with distance $< 20$ km and correlation value $r > 0.80$ (Sibson, 1981).

Outlier values are defined as observations that deviate significantly from the majority of observations, and they are suspected of not being generated by the same mechanisms as the rest of observations (Posio et al., 2008; Hawkins, 1980). Such observations will negatively affect the compatibility and homogeneity of the data, leading to erroneous and inaccurate results (Osborne and Overbay, 2004). For this study, the thresholds of an outlier were defined within the range of ±4 standard deviations (Athar, 2014; Zhang et al., 2005). After that, the comparison with neighboring stations (spatial coherency) was also used to determine whether an outlier was the result of natural factors or systematic errors (Sutarya and Mahendra, 2014). All unreasonable outliers were manually edited by replacing them with monthly averages. Finally, a total of 28 daily maximum and minimum temperature series for the period 1987-2016 fulfilled all the selected criteria, with only two specific stations (Hazeva and Ariel) covering a shorter period (1988-2016 and 1990-2016, respectively).

Mean annual maximum and minimum temperatures vary from 21.0 to 31.6ºC for the maximum, and from 9.8 to 22.4ºC for the minimum, depending on the elevation and distance from the Mediterranean Sea. Table 1 indicates the main characteristics (names, coordinates, availability data, missing values and outliers) of the selected stations used for calculating extreme indices, and their locations are shown in Figure 1. It is noticed that the percent of missing values detected in the whole data set was only 1.3% and the total of outliers detected was lower than 0.01% (Table 1).

Monthly time series from these 28 stations were tested for homogeneity. A homogeneous climate time series is defined as one where variations are caused only by variation in the weather and climate (Conrad and Pollak, 1950; Aguilar et al., 2003). Long climatological time series often contain variations because of non-climatic factors, such as methods of preliminary data treatment, changes in measuring methods, location and in the vicinity of the stations (e.g., urbanization, vegetation) (Peterson et al., 1998; Klok and Klein Tank, 2009; Štěpánek et al., 2013).

For the homogeneity analysis, the method used in this study followed the approach proposed by Wijngaard et al. (2003). Four absolute tests were applied at 5% significance level to evaluate the homogeneity of time series: the Pettitt test (Pettitt, 1979), the standard normal homogeneity test (SNHT) for a single break (Alexandersson, 1986), the Buishand range test (Buishand, 1982) and the von Neumann ratio test (von Neumann, 1941). Time series were labeled as ‘useful’ if one or zero tests rejected the null hypothesis. If two tests rejected the null hypothesis, it was labeled as ‘doubtful’ and as ‘suspect’ if three or four tests rejected the null hypothesis. In addition, time series plotting and comparison (spatial coherence) with neighboring stations (distance $< 20$ km, $r > 0.8$) were used in some cases to evaluate the change points. The results indicated that the monthly time series of January, April, May, October, November and December did not exhibited change points for all stations for the maximum temperature. Additionally, the monthly time series of January, October, November and December exhibited very few change points (less than 5 stations) for minimum temperature. The monthly time series of March, July and August exhibited significant change points for the most stations in the years 2000 and 1997 both for maximum and minimum temperatures. Furthermore, the neighboring stations exhibited the same change points in many cases, for example in some northern stations (stations numbered as 19, 20, 21, and 22, in Fig. 1). However, note that many studies have indicated the relationship between the strong El Niño year 1997/1998 and the change points in temperature series during this year (Tanarhte et al., 2015; Athar, 2014; AlSarmi and Washington, 2011). Therefore, these changes were not considered as inhomogeneities. Finally, all non-climatic change points in the monthly maximum and minimum time series were adjusted before the change points using the software package AnClim V. 5.025.

## 2.2 Extreme temperature indices

In this study, 16 extreme indices derived from daily maximum and minimum temperatures were selected from a total of 27 temperature and precipitation indices recommended by the Expert Team on Climate Change Detection Monitoring and Indices (ETCCDMI) (Folland et al., 1999; Peterson et al., 2001). These indices can be divided into 4 categories (Fonseca et al., 2016; Scorzini et al., 2018): 1) Absolute extreme temperature (TXx, TXn, TNx, TNn and DTR); 2) Percentile-based extreme temperature (TX90p, TX10p, TN90p and TN10p); 3) Duration-based (WSDI and CSDI); and 4) Fixed threshold (SU25/30, TR20/25 and FD). For this study, the thresholds 30ºC and 25ºC were applied for very summer days (SU30) and very tropical nights (TR25) indices, respectively, based on the long-term summer averages for Tmax and Tmin and in a detailed review of the literature for the surrounding regions such as the Arabian Peninsula and Saudi Arabia (Attada et al., 2019; Athar, 2014), Iran (Rahimzadeh et al., 2009; Rahimi and Hejabi, 2018) or Turkey (Erlat and Türkeş, 2013). A brief description and definition of each index is given in Table 2, and further details are available in Zhang et al. (2011) and http://etccdi.pacificclimate.org/list_27_indices.shtml.

Extreme indices were performed using the software package RClimDex v1.0 (Zhang and Yang, 2004) developed at the Climate Research Branch of the Meteorological Service of Canada. The software and documentation are available at http://etccdi.pacificclimate.org. Such software performs the calculations using daily data and provides monthly and annual data of the indices. In addition, it includes a data quality control, which was useful to construct the time series used in this study. All indices were computed at annual time scale, and at seasonal scale for absolute and percentile indices as well as for SU25/30 and TR20/25 indices. Days from December 1$^{st}$ to the end of February were considered for winter (win),

March-April-May for spring (spr), June-July-August for summer (sum) and, from September to November for autumn (aut). Additionally, the annual indices from the 28 stations were spatially averaged for the whole study area.

### 2.3 Trend detection method and atmospheric circulation patterns

Trends in the selected indices were calculated, for each station at annual and seasonal time scales, and averaged over the whole area at annual scale, by Sen's slope estimator (Sen, 1968) meanwhile the Mann-Kendall non-parametric test (MK test, Mann, 1945; Kendall, 1975) was applied to evaluate trend significance. MK test is less sensitive to the non-normality of a distribution and less affected by outliers in the series (Zhang et al., 2000). However, there are some assumptions associated with this test such as the need for statistically independent data (Razavi et al., 2016). For this, the presence of serial correlation was examined prior to application of MK test, and the pre-whitening approach was used to remove the correlation (von Storch, 1995). The statistical significance of the trends was assessed at 0.1, 0.05, 0.01 and 0.001 levels.

Additionally, the monthly values of seven teleconnection indices, NAO, EA, EA/WR, MOI, WEMO, NCPI, and ENSO for the period 1987–2016 were collected from the Climate Prediction Center of the National Oceanic and Atmospheric Administration (http://www.cpc.ncep.noaa.gov/data/teledoc/telecontents.shtml), from Climatic Research Unit of the University of Norwich (https://crudata.uea.ac.uk/cru/data/moi/) for MOI and NCPI, and from the Group of climatology of the University of Barcelona (http://www.ub.edu/gc/en/2016/06/08/wemo/) for WEMO. These monthly values were averaged to obtain seasonal and annual values. Afterward, their influence on the extreme temperature indices was examined by mean of Pearson correlation as in other studies (Unkašević and Tošić, 2013; Efthymiadis et al., 2011; Popov et al., 2018) based on detrended series for each station. The statistical significance of the correlations was assessed at the 5% level.

## 3. Results

### 3.1 Annual trends of extreme temperature indices averaged over the study area

Figure 2 depicts the time series of anomalies for the annual extreme temperature indices averaged over the whole study area during the period 1987-2016, and Table 3 shows their temporal trends. Although the absolute extreme indices (TXx, TNx, TXn and TNn) showed increasing trends by 0.17, 0.40, 0.05 and 0.18$^0$C/decade, respectively, only TNx resulted significant at the 90% confidence level (Table 3). It is also noted that, meanwhile hot absolute extreme indices (TXx, TNx) were found to increase with rates higher than those of their corresponding cold extreme indices (TXn, TNn), the hot percentile TX90p and TN90p indices increase with rates lower than the decreasing rates of the TX10p and TN10p indices. Furthermore, in all absolute and percentile-based extreme temperature indices the increasing trends related to minimum temperature (night-time) indices were higher than those of maximum temperature (daytime) indices (Table 3).

Figures 2c and 2d depict the percentile-based extreme temperature indices TX90p, TX10p, TN90p and TN10p, where the temporal behaviors for TX10p and TN10p clearly pointed to a decreasing trend along the period 1987-2016. Increasing trends by 2.20 and 3.17 days/decade were found, significant at the 99% confidence level, for TX90p and TN90p, respectively; while TX10p and TN10p showed non-significant decreasing trends by -3.94 and -5.52 days/decade, respectively. These results strongly confirmed the general tendency toward the warming over the study area.

The tendency towards warming in the analyzed region was also confirmed by the duration and fixed threshold extreme temperature indices SU25/30 (summer and very summer days), TR20/25 (tropical and very tropical nights), WSDI (warm spell duration) and CSDI (cold spell duration) (Figures 2e, 2f and 2g). The temporal behavior for SU25/30 and TR20/25 indices showed increasing trends along the period 1987-2016. The SU25 and TR20 indices displayed significant increasing trends at 0.001 significance level by 8.15 and 14.4 days/decade, respectively. Moreover, very summer days (SU30) and very tropical nights (TR25) indices exhibited significant increasing trends at the 95% and 99.9% confidence levels by 7.10 and 3.74 days/decade, respectively. Furthermore, WSDI index showed significant increasing trend at the 95% confidence level by 0.63 days/decade, whereas the CSDI index had a significant decreasing trend at the 99% confidence level by -1.79 day/decade (Table 3).

## 3.2 Annual trends of extreme temperature indices at local scale

### 3.2.1 Absolute extreme temperature indices

Figure 3 and Table 4 show the results of the trend analysis for the absolute extreme temperature indices, at annual scale. In general, most stations in the central and southern regions exhibited increasing trends for all absolute indices, although except for TNx, they did not show significant increasing trends. The hot absolute extremes (TXx and TNx) were found to increase at more stations compared to cold extremes (TXn and TNn). In addition, the number of locations that were affected by increasing trends in the minimum temperature indices (TNx and TNn) was larger than those of the maximum temperature indices (TXx and TXn). In detail, maximum values of minimum temperatures (TNx) intensively exhibited increasing trends in 89% (25 stations) of total stations, and 36% (10 stations) with significant increasing trends by values ranging between 0.45 and 1.14ºC/decade. High and significant increasing trends (up to 0.68ºC/decade) were mainly distributed over the Palestinian Coastal Plain regions (Lod Airport, Bet Dagan and En Hahoresh stations) and at some stations in the north (Kefar Blum and Elon stations) and in the south (Elat and Hakfar Hayarok).

The results did not detect any significant decreasing trend in the five absolute indices, except for one station (Hazeva) in ETR index.

### 3.2.2 Percentile-based extreme temperature indices

The results shown in Table 4 and Figure 4 clearly indicated homogeneous patterns of the increasing/decreasing trends for all percentile-based extreme temperature indices. The results also confirmed the tendency toward warming over the study area where all locations exhibited decreasing trends in their cold extreme (TX10p and TN10p) indices whereas they showed increasing trends in their hot extreme (TX90p and TX10p) indices. In this context, 100% of total stations exhibited increasing trends in their TX90p and TN90p, with 89% (25 stations) and 93% (26 stations), respectively, showing significant trends with values up to 2.74 and 3.20 days/decade, respectively. The high band of increasing trends higher than 2.39 days/decade affected more locations for TN90p than those for TX90p, such as some stations in the Palestinian Coastal Plain region (En Hahoresh, Galed, Lod Airport and Bet Dagan) (Figure 4). Coherently, TX10p and TN10p indices displayed decreasing trends in 100% of total stations but with a smaller number of significant stations, only 1 and 3 stations, respectively (Table 4).

### 3.2.3 Duration and fixed threshold extreme temperature indices

The tendency toward a warming was also confirmed by the results shown in Table 4 and Figure 5 for the annual duration and fixed threshold extreme temperature indices, where decreasing trends were not observed for WSDI, SU25/30 and TR20/25 indices, except for Ariel station, although was not significant. In addition, the WSDI index showed increasing trends while the CSDI index showed decreasing trends.

Summer days (SU25) and tropical nights (TR20) indices exhibited very intense and significant increasing trends by 82% and 86% (23 and 24 stations, respectively). All stations that showed significant positive trends in their SU25 index, also showed significant positive trends in their TR20 index. Furthermore, most regions in the study area were affected by a high band of significant increasing trends (between 6.90 and 11.09 days/decade) in both indices (Figure 5).

For very summer days (SU30) and very tropical nights (TR25) indices, the significant increasing trends intensively occurred at more locations for SU30 (57%, 16 stations) than those of TR25 index (28.5%, 8 stations). The very high and significant increasing trends (between 17 and 22 days/decade) covered the north-western regions at Elon and Akko stations for SU30, while the high and significant increasing trends (from 11 to 14 days/decade) found over the south-eastern areas at Elat, Yotvata and Hazeva stations, and the north-eastern areas at Yavne'el and Massada for TR25 index (Figure 5).

The warm spell duration index (WSDI) denoted increasing trends in 52% (13 stations), and only 39% (11 stations) resulted significant at the 95% significance level (Table 4). The positive increasing trends (between 1.26 and 2.83 days/decade) were mainly distributed over the northern and center regions (Figure 5).

The negative trends for cold spell duration index (CSDI) and frost days (FD) also confirmed the tendency toward the warming, with 28.5% of the locations (8 stations) showing decreasing trends for CSDI index and 21% of them (6 stations) with significant trends. Only one station (Lahav) displayed significant increasing trend by 1.00 day/decade for CSDI. High decreasing trends of CSDI between -1.99 and -2.49 days/decade occurred in the north-western region. For FD index, most stations (57%) showed negative trends, but only 1 station (Akko station) resulted significant with a value of -0.12 days/decade. In addition, for this latter index, only few isolated stations presented positive trends although non-significant.

### 3.3 Seasonal trends of extreme temperature indices at local scale

#### 3.3.1 Absolute extreme temperature indices

Results presented in Table 5 and Figures 6 and 7 did not reveal any significant decreasing trend for all the seasonal absolute extreme temperature indices. Furthermore, warming trends affected more locations than the decreasing trends for all the seasonal absolute extreme temperature indices, except for TXx-spring and TXx-autumn (Table 5 and Figure 6), when non-significant decreasing trends affected more locations, mainly distributed in the northern region.

The analysis for frequency and intensity of trends also revealed that the total number of locations that was affected by significant/non-significant warming trends for the minimum temperature indices (TNx and TNn, Figure 7) were more than those for the maximum temperature indices (TXx and TXn, Figure 6), especially for spring and autumn (Table 5).

More than 96% of total stations showed positive trends for all absolute summer extreme temperature indices (Table 5). Moreover, summer did not show any decreasing trend at all stations and for all indices. It is also interesting to note that it exhibited the highest total of stations with significant warming trends by 50%, 89%, 96% and 89% of total stations respectively for the TXx, TXn, TNn and TNx indices (Table 5). For spring and autumn, they showed dominant warming trends (> 96% of total stations) in the minimum values of the absolute temperature indices (TXn and TNn), with significant trends for more than 43% of total stations, while winter showed dominant occurrence of warming trends for maximum values of temperature indices (TXx and TNx), with significant results for more than 29% of total stations.

Regarding decreasing trends, spring and autumn displayed decreasing trends by 50% of total stations for TXx index. They also displayed decreasing trends by 29% and 11% of total stations for TNx index, respectively. Furthermore, the decreasing trends occurred in TXn-winter and TNn-winter by 25% and 36%, respectively (Table 5). The coherent spatial of decreasing trends, although nonsignificant, were found in some cases such as for TXx-spring in the northern regions, TNx-spring in the northern and central areas of the West bank, and TXx-autumn in the regions that extended from central West bank to the most north-eastern regions for the northern regions (Figures 5 and 6).

The analysis of the intensity of warming trends revealed that summer significantly showed the highest warming trends between 0.98 and 1.36°C/decade for TXx index and spatially covered the regions extended eastward and northward of the Gaza strip. It is also exhibited a significant warming trend between 0.6-1.0°C/decade in the north-eastern region. In addition, spring and autumn exhibited warming trends in TXn index stronger and more pronounced than their trends in TXx index. The highest rates of warming trends occurred in autumn for TXn index, from 0.90 to 1.86°C/decade, distributed over the northern regions of the study area (Figure 6). Furthermore, the high significant increasing trends for summer (from 1.0 to 1.22°C/decade) covered the north-eastern regions and the upper areas of the Negev desert.

For the minimum temperature indices (TNx and TNn), the results shown in Figure 7 generally indicated that spring and autumn exhibited more intense and positive trends for (TNx) index than their (TXx) index. Similar conditions were also found in summer, especially for the northern and south-eastern regions. Moreover, summer showed more intense increasing trends for TNn index than TXn index, especially for all coastal regions that extends from Gaza strip in the south to Akko in the north (with rates from 0.98 to 1.47°C/decade). For TNx index, the highest and significant warming trends, in the interval 1.11-1.23°C/decade, appeared in summer, and spatially distributed in the Palestinian Coastal Plain regions (Beit Dagan, Lod Airport and En Hahoresh stations) as well as at Kefar Hayarok and Harashim stations in the north. For TNn index, autumn showed high increasing trends in the northern regions with some locations reaching 2.33°C/decade.

### 3.3.2 Percentile-based extreme temperature indices

In general, all seasons showed significant increasing trends in the TN90p index (Figure 9) for more locations and with higher intensity than the TX90p index (Figure 8), except for winter. For the TX90p index, winter and summer significantly displayed higher frequency and intensity of the increasing trends compared with those for spring and autumn, when 61%, 32%, 75% and 3.5% of total stations exhibited significant increasing trend respectively for winter, spring, summer and autumn (Table 5). The strongest significant increasing trends (from 3.74 to 4.27 days/decade) covered the most northern and southern regions in winter and they covered all regions, except very few locations in summer (Figure 8). In spring, the area extended eastward and northward of the Gaza strip showed a significant increasing trend in their TX90p index by 2.70 days/decade on average.

For TN90p index, 43%, 61%, 86% and 32% of total stations had significant increasing trends for winter, spring, summer and autumn, respectively (Table 5). The strongest significant positive trends were in summer (by 4.79-5.31 days/decade) and spatially covered all the study area. For the rest of seasons, the increasing trends were, in general, less intense (< 3.22 days/decade) for all regions (Figure 9).

Regarding TX10p and TN10p indices, all seasons had decreasing trends for all locations, except a total of six stations (three in TX10p-summer and three in TN10p-winter) which showed very weak and non-significant increasing trends. The results also indicated that winter did not show significant decreasing trends in both indices and for all locations. Contrariwise, autumn showed the highest occurrence and intensity of decreasing trends, significant for the 86% of the total number of stations for TX10p index and the 57% for TN10p index (Table 5). For TX10p-autumn index (Figure 8), the highest significant decreasing trends (between -2.78 and -5.36 days/decade) covered the entire study region. For TN10p-autumn (Figure 9) the intensity of decreasing trends reached –values even higher in general.

Spring, showed significant decreasing trends at more locations and with higher intensity in TN10p index than TX10p index (50% and 7% of total stations, respectively, Table 5). For TN10p-spring, most locations showed significant decreasing trend (from -2.52 to -3.57 days/decade). In addition, the significant decreasing trends showed an generalized pattern in TX10p-summer and TN10p-summer, but only with five significant results (Figures 8 and 9).

### 3.3.3 Duration and fixed threshold extreme temperature indices

The results presented in Figure 10 and Table 5 indicated 82% of total stations (23 stations) remarkably showed significant and positive trends for TR20 index (between 4.39 and 7.53 days/decade) in summer. The regions extended eastward and northward of the Gaza strip had the highest positive trends in summer, with rates from 6.28 to 7.53 days/decade. Furthermore, 64% and 39% of total stations had significant increasing trends respectively in autumn and spring. For autumn, the trends ranged between 3.14 and 3.77 days/decade, whereas they were lower than 1.43 days/decade for most locations in spring. In addition, the northern regions showed decreasing trends for TR20-spring by -0.62 days/decade (Table 5, Figure 10).

For SU25 index, summer did not exhibit remarkably increasing or decreasing trends in 89% of total stations, whereas winter showed significant increasing trends in 71% of the cases (20 stations), autumn in 14% (4 stations) and spring in 75% (21 stations). Spring had the highest significant and generalized increasing trends, up to 4.39 days/decade, which are mainly distributed over the whole area (Figure 10).

For autumn and summer, the frequency and intensity of increasing trends related to the minimum temperature (TR20) were greater than those of maximum temperature (SU25) at all locations. Contrary, the frequency and intensity of increasing trends for winter and spring were greater for SU25 than for TR20 at all locations.

For very extreme threshold indices, the increasing trends for SU30-spring were more pronounced than those for TR25-spring for all regions, except at Ariel and Jerusalem stations which exhibited non-significant decreasing trends (Figure 11). In this context, for TR25-spring only 6 stations showed increasing trends with 2 stations of them being significant, while 75% (21 stations) showed increasing trends for SU30-spring index with 33.4% (7 stations) of them being significant. Spatially, the south-eastern regions along the border with Jordan exhibited increasing trends for both indices, but with higher rates (from 2.52 to 3.77 days/decade) for SU30 index (Figure 11).

For winter, the results did not detect any trend for TR25 and SU30, except at Elat station which exhibited a significant increasing trend for SU30 by 0.43 days/decade (Figure 11).

In summer, TR25 index significantly showed increasing trends at more locations (53% of total stations) than those for SU30 index (28% of total stations), whereas SU30 generally showed higher intensity of increasing trends than TR25 in most locations, especially in the Palestinian coastal Plain areas and some locations in the northern regions. Most stations in the Palestinian coastal plain showed significant increasing trends for TR25 index (from 0.64 to 1.26 days/decade). These regions also showed high increasing trends, although nonsignificant, for SU30 index (from 5.65 to 6.28 days/decade). It is also noticed that the southern regions showed higher rates of increasing trends in TR25 index (from 5.02 to 6.28 days/decade).

For autumn, TR25 index did not show increasing/decreasing trends in 71% of total stations. On the other hand, SU30 index exhibited increasing trends in 93% of the cases (26 stations), but only 4 stations of them being significant at the 95% confidence level.

### 3.4 Extreme temperature indices and teleconnection patterns

In this section, the relationships between the extreme temperature indices and the selected large-scale circulation patterns were examined, with the aim to identify whether any specific circulation pattern could have some influence on the occurrence of the temperature extremes over the study area. Tables 6 and 7 summarize the number of stations with significant correlations between the extreme temperature indices and the NAO, EA, EA/WR, WEMO, MOI, ENSO and NCPI teleconnection indices at annual and seasonal scales, respectively. The results listed in Tables 6 and 7 generally demonstrated the NCPI is the main driver of the extreme temperature indices over the study area. For the sake of saving space, we present here only the correlation coefficients between NCPI and the extreme temperature indices at annual and seasonal scales (Figures 12 and 13, respectively). Moreover, we provide four supported files (S1, S2, S3 and S4) of the main correlation coefficients for EA/WR, NAO and ENSO patterns. In addition, Figure 14 shows the temporal variability of the NCP mode standardized of the index with some standardized temperature indices time series.

### 3.4.1 Annual scale

According to Table 6, the NCP pattern intensively showed the most significant influence on TXx, TXn, TX90p, TX10p, TN90p, TN10p, WSDI, SU25, SU30, CSDI, TR20 and TR25 indices by 28.6%, 57%, 93%, 82%, 86%, 75%, 71.5%, 46%, 64.3%, 71.4%, and 60.7% of total stations, respectively,. Moreover, all correlation coefficients between the NCPI and all extreme temperature indices were negative, except for TN10p and for TX10p, CSDI and FD indices in some locations.

The highest negative correlation coefficients (between -0.56 and -0.63) were found with the TX90p, TN90p and WSDI indices and spatially covered the most southern regions, the upper parts of the Negev desert as well as the northern regions (Figure 12). High correlation coefficients (between -0.43 and -0.56) covered the northeastern regions and Jerusalem district for TXx, while for TXn, correlations up to -0.50, covered the southern areas of the Palestinian Coastal Plain and northeastern regions. For TN10p, most stations showed significant and positive correlations (up to 0.47) (Figure 12).

Regarding the other patterns, the EA/WR represented the second most important pattern affecting the extreme temperature indices over the study area (Table 6). Its influence mainly occurred over SU25 index by 86% of total stations which significantly presented negative correlations with it. High and significant correlations (up to -0.63) covered the Palestinian Coastal Plain, Beer Sheva and the northeastern districts (Figure S1 in the supplementary material). In addition, it showed significant correlations, although lower, with TX10p (43% of total stations), TN10p (39%), SU30 (43%), WSDI (43%) and CSDI (46%) indices at more locations than other patterns (Table 6).

The main influence of the EA pattern was on TXx, TNx and TX90p indices by 28.5%, 39% and 28.5% of the total stations, respectively, and with correlation coefficients ranging from 0.31 to 0.48 (not shown). The NAO pattern also exhibited significant and positive correlations with TNx by 36% of total stations, with correlation coefficients ranging from 0.32 to 0.48. Furthermore, the NAO index significantly displayed negative correlations with TX90p and TN90p by 61% and 50% of total stations, respectively. For TX90p, high correlation between -0.41 and -0.61 occurred in the most southern region as well as in the eastward and northward of the Gaza strip. For TN90p, the highest significant correlations were reached at the centre region, up to -0.72. (Figure S2 in the supplementary material).

The significant correlations were very weak between MOI, WEMO and ENSO patterns and all extremes indices, except the positive correlation with 32% of total stations between MOI and WSDI, 32% of total stations between WEMO and TN10p and 50% of total stations between ENSO and TN10p

**3.3.2 Seasonal scale**

Table 7 summarizes the number of stations with significant correlations between the extreme temperature indices and the teleconnection patterns at seasonal scales. The results also indicated the NCP pattern was the main driver of extreme temperature indices in winter for 5 out of 12 indices, and in autumn for 8 out of 12 indices, while EA/WR was the main driver of extreme indices in spring for 5 out of 12 indices. In summer, the ENSO pattern showed a big influence on 5 extreme temperature indices. In general, the main influence seems be exerted by the NCP pattern, concentrated over the percentile extreme temperature indices for winter and autumn (for all percentile extremes) and in summer (for TX90p and TN90p).

In winter, the NCP pattern intensively presented the highest correlation coefficients, compared with the other patterns, on TXn, TNn, TX90p, TN90p, TX10p and TN10p indices with 100%, 86%, 75%, 61%, 79%, and 96% of total stations, respectively (Table 7, Figure 13). High and negative correlations (between -0.52 and -0.72) covered mostly the study area for TXn, TNn, TX90p and they concentrated in the north-eastern regions for TN90p. For positive correlations with TX10p and TN10p, very high and significant values (from 0.63 to 0.72) spatially covered all the study area for both indices. Figure 14b illustrates the strong positive correlation between the temporal variability of the standardized NCPI and TN10p index averaged for all locations along the period 1986-2016 in winter.

In spring, the NCPI also had the highest influence on TNn with 39% of total stations showing significant correlations as well as it exhibited high occurrence of significant correlations with TXn, TX90p and SU25/30 with 68%, 32% and 68/50%, respectively (Table 7). The NCPI in summer was the main controller for TX90p, TN90p and TR25 indices respectively by 61%, 50% and 32% of total stations.

In autumn, the NCP pattern showed high influence on 8 out of 12 indices, TNn, TX90p, TN90p, TX10p, TN10p, SU25/30 and TR20 indices, respectively with 79%, 96%, 79%, 32%, 61%, 100/68% and 75% of total stations with significant correlations. It also showed high occurrence of significant correlations with TXn and TNx indices with 61% and 39% of total stations, respectively (Table 7). For TX90p and TN90p, high correlation coefficients (from -0.59-to -0.67) spatially covered the complete study region (Figure 13). A similar pattern was found for TNn index, but with lower correlations. For TN10p index The correlation pattern showed positive values extended over all the region. The temporal variability of the standardized NCPI and TX90p (Figure 14a) and SU25 (Figure 14c) indices showed the strong negative correlations along the period 1986-2016.

Regarding EA/WR pattern, it resulted the main controller for TXx, TXn, TN90p and TR20/25 indices for 96%, 82%, 36% and 50/29% of total stations in spring (Table 7). Spatially, high correlation coefficients (from -0.42- to -0.50) were found in the Palestinian Coastal Plain and north-western areas for TXn index, as well as in all regions of the study area for TXx index (Figure S3 in the supplementary material). The EA/WR pattern also displayed a high frequency of significant correlations with SU25/30 indices by 79/50% of total stations (Table 7).

In summer, the EA/WR pattern was the main controller for all absolute extreme temperature indices for 53%, 61%, 100% and 89% of total stations for TXx, TNx, TXn and TNn, respectively. The highest correlation coefficients were found for TXn and TNn indices, between -0.41 and -0.61 (Figure S3 in the supplementary material).

On the other hand, in autumn, the EA/WR pattern displayed the highest influence on TXx and TNx for 96% and 75% of total stations. All regions showed correlations coefficients (between -0.41 and -0.50) for TXx, except for the northeastern region where the correlations were lower (Figure S3 in the supplementary material).

Compared with the other patterns, the NAO showed the highest effect on SU25/30 for 71/68% of total stations in winter (Table 7). Strong correlation coefficients (between -0.51 and -0.72) were found in the area extended from the centre to the south of the region for SU25index. A similar pattern was also found for SU30, but showing lower correlations (Figure S4A in the supplementary material). In autumn, NAO pattern had the highest significant correlations with TXn for 89% (Table 7).

It is noticed that the ENSO pattern showed the highest influence on TX10p and TN10p indices for 68% and 61% of total stations, respectively, during spring and summer. Spatially, high and significant correlations between 0.42 and 0.62, were expanded over the study region (Figure S4B in the supplementary material).

## 4. Conclusions and discussion

Topic such as spatio-temporal analysis for extreme temperature indices over Israel and Palestine in the Levant region were not fully explored by previous studies, and it is required further attention. Thus, the main aim of this study was to provide a comprehensive analysis of this topic using more stations and for a recent period of time. In this study, trends for 16 extreme temperature indices computed from daily data of 28 stations homogeneously distributed over the territory were spatially and temporally analyzed at annual and seasonal scales, for the period 1987-2016. In addition, the relationships between these extreme indices and large-scale circulation patterns (NAO, EA, EA/WR, WEMO, MOI, NCPI and ENSO) were examined for each station. The main conclusions can be summarized as follow:

1. The analysis of 16 extreme temperature indices revealed a dominant warming tendency for the last three decades over the study area. Extremes related to minimum night-time temperature denoted more intense trends compared to those of maximum day-time temperature indices, at annual and seasonal scales.
2. Regarding the averaged extreme indices over the study area, significant increasing trends for seven extreme temperature indices (TNx, TX90p, TN90p, SU25, SU30, TR20 and TR25) were detected at annual scale, whereas the significant decreasing trends were detected for the CSDI and TN10p indices. These results are consistent with other studies in regions around the study area. Donat et al. (2014) analyzed the changes in extreme temperature and precipitation over the Arab region based on data of 61 stations, and they found increasing trends in the averages of TXx, TX90p, TN90p and WSDI indices by 0.23, °C/decade, 1.6, 2.1 and 0.76 days/decade, respectively. They also found decreasing trends in CSDI, TX10p and TN10p indices by -3.3, -2.2 and -3.2 days/decade, respectively. Over Central Asia, Feng et al. (2018) studied the spatio-temporal variations in extreme temperature, based on data of 108 stations from six countries for 1981-2015, and the results revealed that both TX90p and TN90p experienced significant increasing trends by 1.1 and 1.4 days/decade, respectively as well as TX10p and TN10p showed significant decreasing trends by -1.01 and -0.62 days/decade, respectively. Furthermore, Almazroui et al. (2014) analyzed trends of temperature extremes in Saudi Arabia based on data of 27 stations during the period 1981-2010, and they found significant increasing trends in the TNx, TX90p, TN90p and WSDI indices by 0.7$^{0}$C/year, 16.9, 12.7 and 3 days/decade, respectively. The results also indicated the significant decreasing trends in CSDI,

TX10p and TN10p index by -2.4, -19.4 and -16.3 days/decade. Finally, Erlat and Türkeş (2013) analyzed trends of SU25 and SU30 indices over Turkey based on data of 97 stations during 1976-2010 and the results revealed significant increasing trends by 6.8 and 7.2 days/decade, respectively.

3. At annual and local scale, the analysis of trends for extreme temperature indices revealed significant warming trends in TNx index (36% of total stations) with averages ranging between 0.45 and 1.14ºC/decade. Moreover, the analysis of trends of extreme temperature indices revealed significant increasing trends in WSDI, SU25/30 and TR20/25 indices. For SU25 and TR20, more than 82% of total stations were affected by significant increasing trends with averages ranging between 3.15 and 11.40 days/decade while SU30 and TR25 indices exhibited significant increasing trends in 46% and 29% of total stations. For WSDI index, 39% of total stations were affected by significant increasing trends with averages from 0.24 to 2.96 days/decade. In addition, the percentile-based extreme temperature indices showed very coherent patterns for significant increasing trends with 89% and 92% of total stations respectively exhibited significant increasing trends for TX90p and TN90p indices. Furthermore, TX10p and TN10p indices showed decreasing trends in 100% of total stations. In agreement with these results, studies carried out in surrounding regions provide results in the same line. Rahimzadeh et al. (2009) analyzed the variability of extreme temperature over Iran during 1951-2003, and the results revealed that 46%, 74%, 46%, 44% and 85% of total stations (27 stations) exhibited significant increasing trends in their TNx, TN90p, TX90p, SU25 and TR20 indices, respectively. For TX10p and TN10p indices, Rahimzadeh et al. (2009) also found decreasing trends in more than 55% of total stations. Furthermore, Rahimi and Hejabi (2018) studied extreme temperature indices over Iran during 1960-2014 and they found 88%, 71%, 76%, 61% and 73% of total stations (33 stations) had statistically significant increasing trends in their TNx, TX90p, TN90p, SU25 and TR20 indices whereas TX10p and TN10p indices exhibited significant decreasing trends in 39% and 64% of total stations, respectively. The TNx, TR20 and WSDI indices also showed intensive significant increasing trends in 70%, 74% and 70% of total stations (27 stations) over Saudi Arabia (Almazroui et al., 2014). Finally, Erlat and Türkeş (2013) found 95% and 94% of total stations (97 stations) showed upward trends in their annual averages of SU25 and SU30 indices, respectively over Turkey.

4. At seasonal scale, the results did not indicate significant decreasing trends for any absolute extreme temperature indices. For TX10p-autumn 86% of total stations showed significant decreasing trends as well as 50% and 57% of total stations in their TN10p-spring and TN10p-autumn, respectively. Contrary, at annual scale, the analysis of trends for extreme temperature indices showed intense and broad significant increasing trends in all the absolute extreme temperature indices. For summer, more than 90% of total stations exhibited significantly increasing trends for TXn, TNn and TNx indices with averages ranging between 0.38 and 2.0ºC/decade. For autumn and spring, more than 40% of total stations had significant increasing trends of TXn, TNn indices with averages ranging from 0.55 to 2.33ºC/decade for TXn, and between 0.5 and 1.89˚C/decade for TNn. In winter, 39% of total stations showed significant increasing trends in TNx with averages ranging between 0.56 and 1.0ºC/decade. The intense and coherent increasing trends were found also in TX90p and TN90p indices for 61% and 75% of total stations for TX90p-winter and TX90p-spring, respectively. For TN90p, 61% and 85% of total stations showed significant increasing trends for TN90p-spring and TN90p-summer. These results are in agreement with those obtained in surrounding areas. Islam et al. (2015) analyzed the changes in seasonal temperature extremes over Saudi Arabia during 1981–2010, and they found that summer exhibited significant warming trends for all absolute extreme temperature indices more than those for other seasons by 63% for TXx, 48% for TXn, 59% for TNn and 70% for TNx of total stations (27 stations). The analysis of seasonal trends in frequency of extreme temperature indices revealed significant increasing trends of SU25 index in winter and spring (71% of total stations) with averages ranging between 0.1-5.87 days/decade. In summer and autumn, more than 64% of total stations exhibited significant increasing trends of TR20 index by averages ranging between 1.3 and 5.14 days/decade. TR25 showed significant increasing trend in 54% of total stations for summer with averages ranging 0.04-9.19 days/decade. Note that the warming process in densely populated regions like Jerusalem, Akko and Beer Sheva is expected to be even faster due to the urban heat island, at least partly (Ziv et al., 2005). Many studies indicated the effect of urbanization on the increasing temperature in cities, especially during summer, when urban heat islands are stronger due to a greater storage of heat in urban structures (Fujibe, 2009). Itzhak Ben Shalom et al. (2016) studied the trends of urban warming during the period 1980-2014 in four Israeli cities (Jerusalem, Beer Sheva, Elat and Tel Aviv) to estimate the urbanization effect on the local climate, and they found that the urban minus rural temperature showed a more intense warming in the daytime in the 4 cities.

In the present study, the urban locations of Jerusalem and Akko showed the highest increasing trends above 6.38 days/decade in SU30-summer index, but other urban stations as Elat and Beer Sheva did not show significant increases in this index. Other rural locations, such as Elon and Bet Dagan, also exhibited significant increasing trends in SU30-summer. Moreover, for TNx index in summer almost all the locations showed significant trends with similar values, independently of the urban or rural character. Therefore, from our study we cannot confirm the urban heat effect as the responsible of the warming trends in Levant region. The generalized warming trends could be due to the global warming (Parker, 2004).

5. For the analysis of the influence of teleconnection indices on the extreme temperature over Israel and Palestine, the study revealed the NCP pattern was the main driver of extreme temperature variability over the study area, particularly at annual, winter and autumn scales. It displayed strong influence compared with other patterns on 11 out of 12 indices for the annual extremes, 5 out of 12 indices for winter and 8 out of 12 indices for autumn. In addition, EA/WR and ENSO displayed notably effects on the extreme indices in some cases, especially in spring and summer. The study also revealed the main influence of the NCPI generally concentrated in the percentile extreme temperature indices for winter (all percentile extremes), summer (TX90p and TN90p) and autumn (all percentile extremes). For EA/WR, it represented the second most important pattern affecting the extreme temperature indices over the study area, and its main influence occurred on the absolute extreme temperature indices for winter (TNx), summer (all absolute extremes), autumn (TXx and TNx) and spring (TXx and TXn). At annual scale, strong significant correlations were observed between the NCPI and TX90p, TX10p, TN90p, TN10p, WSDI and TR20 indices respectively, by 93%, 82%, 86%, 75%, 71.5%, and 71.4% of total stations. In winter and autumn, the NCPI broadly and intensively exhibited significant correlations with TXn-winter, TNn-winter, TN10p-winter, TX90p-autumn and SU25-autumn respectively by 100%, 86%, 96%, 96% and 100% of total stations, respectively. On the other hand, EA/WR had the highest effect compared with other patterns on TXx-spring, TXn-spring, TXn-summer, TNn-summer by 96%, 82%, 100% and 89% of total stations. For ENSO pattern, it exhibited the highest effect in summer on TX10p, TN10p, SU25/30 and TR20 by 61%, 53.5%, 18/36% and 36%, respectively. Other teleconnection patterns (NAO, EA, WEMO and MOI) did not show significant correlations, except in very few cases such as WEMO in spring with TNx (36% of total stations), TX90p (36% of total stations) and SU25 (86% of total stations).

These results are in accordance with those carried out in surrounding areas. Kutiel and Benaroch (2002) defined the North Sea-Caspian Pattern (NCP) as atmospheric teleconnection in the 500 hPa geopotential height between two regions centred in [0˚-10˚E, 55˚N] for its north-western pole and between [50˚E-60˚E, 45˚ N] for its south-eastern pole. They concluded that the main impact of this teleconnection should be exhibited over the Balkans and the eastern Mediterranean basin mainly in autumn, winter and spring, and is less frequent in summer. Furthermore, Kutiel et al. (2002) used data of NCP, monthly mean air temperature and monthly total rainfall from 33 stations across Greece, Turkey and Israel for the period 1958-1998 to analyze the implication of the NCP on the regional climate of the eastern Mediterranean basin. Their results confirmed that the NCP was the main atmospheric teleconnection affecting the climate of the Balkan, the Anatolia Peninsula and the Middle East region. The positive phase of this pattern is associated with below normal temperature, while the negative phase temperature is related with above normal temperature. In addition, they found the impact of NCP on air temperature was more severe in Turkey, due to its vicinity to one of the poles of the NCP, and its impact was more pronounced in the mountainous inland regions for Israel. In addition, the results found there was more rainfall during the positive phase of the NCP over all regions in Israel. Ghasemi and Khalili (2008) analysed the effect of the NCP on winter temperatures in Iran based on data of 31 stations for 1958-2000, and they found negative and significant correlations between the NCPI and minimum, maximum and mean winter temperature for 90%, 87% and 97% of total stations, respectively. Their results shown that the positive NCP is associate with enhanced precipitation and cloudy conditions, which causes below normal temperatures over Iran. As noted by Brunetti and Kutiel (2011), during the positive phase of NCPI in winter, the anomaly circulation is mainly northerly over major parts of Eastern Europe, the Black and the Caspian seas and the Eastern Mediterranean area, leading a considerable decrease in temperatures compared to the normal conditions. The opposite is true during the negative phase of NCP.

For the impact of EA/WR pattern, Baltacı et al. (2018) analyzed relationships between teleconnection patterns and Turkish climatic extremes based on data of 94 stations for 1965-2014, and the results indicated that EA/WR pattern intensively exhibited negative and significant correlations with TNx, TNn and TXn indices in winter, stronger than those from the NAO and EA patterns. These authors established that the positive phase of EA/WR is associated with a prevalence of north flow over the Eastern Mediterranean area, which can explain the sign of the correlations found. Other studies have been also identified the influence of the EA/WR pattern on the climate of the region, but they do not use the temperature (Yosef et al., 2009; Krichak and Alpert, 2005). Further studies are needed to understand the underlying physical processes explaining the links to large-scale circulation patterns over the study area.

The knowledge and understanding the changes in extreme temperature indices over Levant region can be beneficial for many sectors, such as in projects related to water and energy supply. It is also important for policy makers in socio-economic planning, tourism, ecosystems and agriculture.

**Acknowledgments**

The Spanish Ministry of Economy and Competitiveness, with additional support from the European Community Funds (FEDER), projects CGL2013-48539-R and CGL2017-89836-R, financed this study. We thank to the anonymous reviewers for their suggestions and comments.

Figure captions

Figure 1. Study area overview and spatial distribution of the stations considered in this study. Names of stations are mentioned in Table 1.

Figure 2. Time series of anomalies for the annual extreme temperature indices averaged over the study area during the period 1987–2016.

Figure 3. Spatial distribution of trends (ºC/decade) for absolute extreme temperature indices.

Figure 4. Spatial distribution of trends (days/decade) for annual percentile-based extreme temperature indices.

Figure 5. Spatial distribution of trends (days/decade) for annual duration and fixed threshold extreme temperature indices.

Figure 6. Spatial distribution of trends (ºC/decade) for seasonal absolute extreme maximum temperature indices.

Figure 7. Spatial distribution of trends (ºC/decade) for seasonal absolute extreme minimum temperature indices.

Figure 8. Spatial distribution of trends (days/decade) for seasonal percentile-based extreme temperature indices.

Figure 9. Spatial distribution of trends (days/decade) for seasonal percentile-based extreme temperature indices.

Figure 10. Spatial distribution of trends (days/decade) for duration and fixed threshold extreme temperature indices.

Figure 11. Spatial distribution of trends (days/decade) for duration and fixed threshold extreme temperature indices.

Figure 12. Spatial distribution of Pearson correlation coefficients between the NCP index and the extreme temperature indices at annual scale.

Figure 13. Spatial distribution of Pearson correlation coefficients between the NCP index and the extreme temperature indices at seasonal scale.

Figure 14. Temporal variability of the standardized atmospheric/oceanic mode index with some standardized temperature indices time series.

Figure S1. Spatial distribution of Pearson correlation coefficients between the EA/WR index and the extreme temperature indices at annual scale.

Figure S2. Spatial distribution of Pearson correlation coefficients between the NAO index and the TX90p and TN90p indices at annual scale.

Figure S3. Spatial distribution of Pearson correlation coefficients between the EA/WR index and the extreme temperature indices at seasonal scale.

Figure S4. Spatial distribution of Pearson correlation coefficients between A) the NAO and SU25/30 in winter, and B) the ENSO and TX10p and TN10p in spring and summer.

Table captions

Table 1. Meteorological stations used in this study.

Table 2. Description of extreme temperature indices used in this study.

Table 3. Trends for the annual extreme temperature indices averaged over the study area. Symbols ***, **, * and +, indicate a significant trend at $\alpha = 0.001$, $\alpha = 0.01$, $\alpha = 0.05$ and $\alpha = 0.1$ level, respectively.

Table 4. Number of stations showing positive and negative trends for each extreme temperature index at annual scale. In bracket the number of stations showing significant trends at the 95% confidence level. Right column shows the percentage of stations with significant increasing/decreasing trends.

Table 5. Number of stations with positive and negative trends for each extreme temperature index at seasonal scale. In bracket the number of stations with significant positive or negative trends at the 95% confidence level.

Table 6. Number of stations with significant positive or negative correlations between extreme temperature and teleconnection indices at annual scale. Only significant results at the 95% confidence level are shown.

Table 7. Number of stations with significant correlations between extreme temperature and teleconnection indices, at seasonal scale. Only significant results at the 95% confidence level are shown.

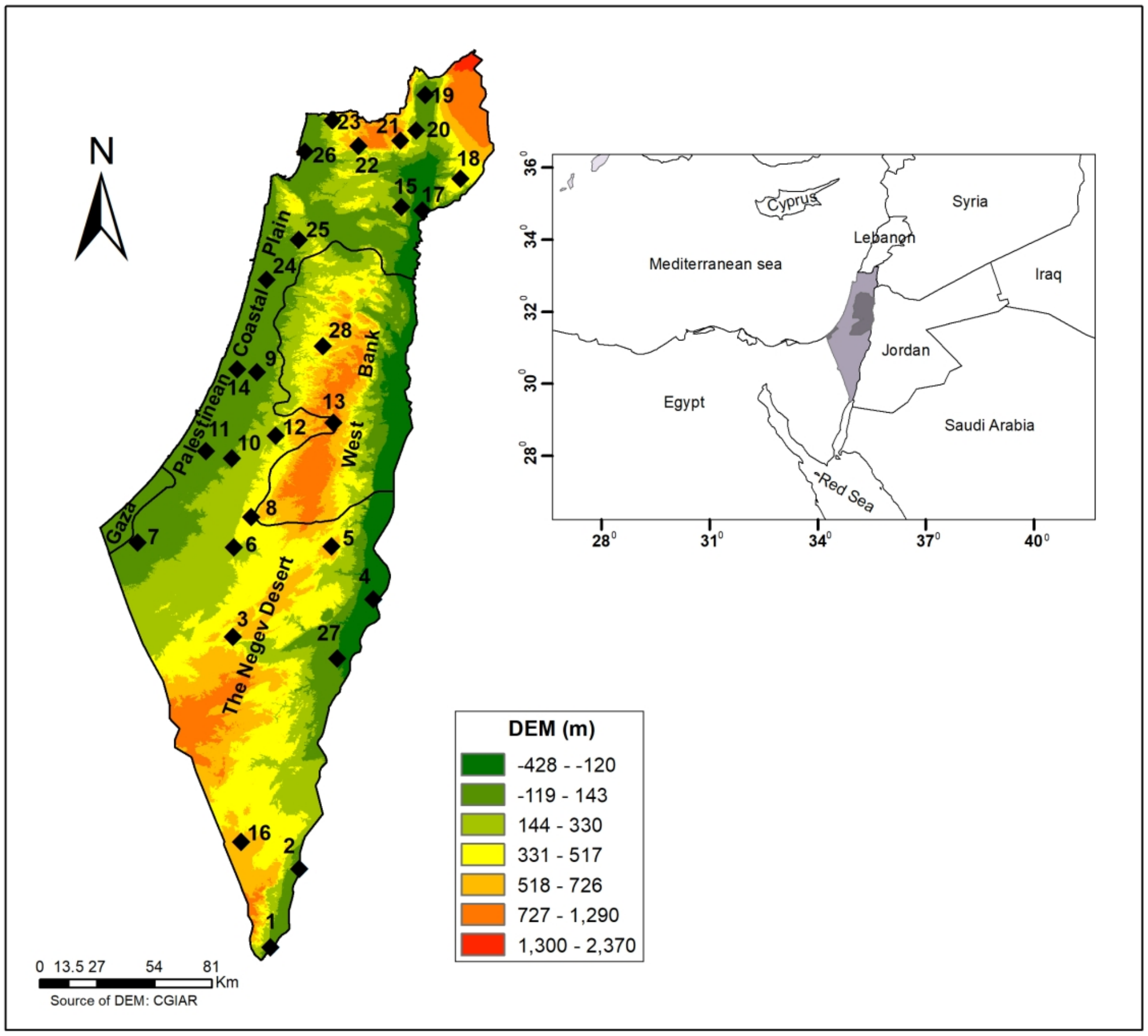

Figure 1. Study area overview and spatial distribution of the stations considered in this study. Names of stations are mentioned in Table 1.

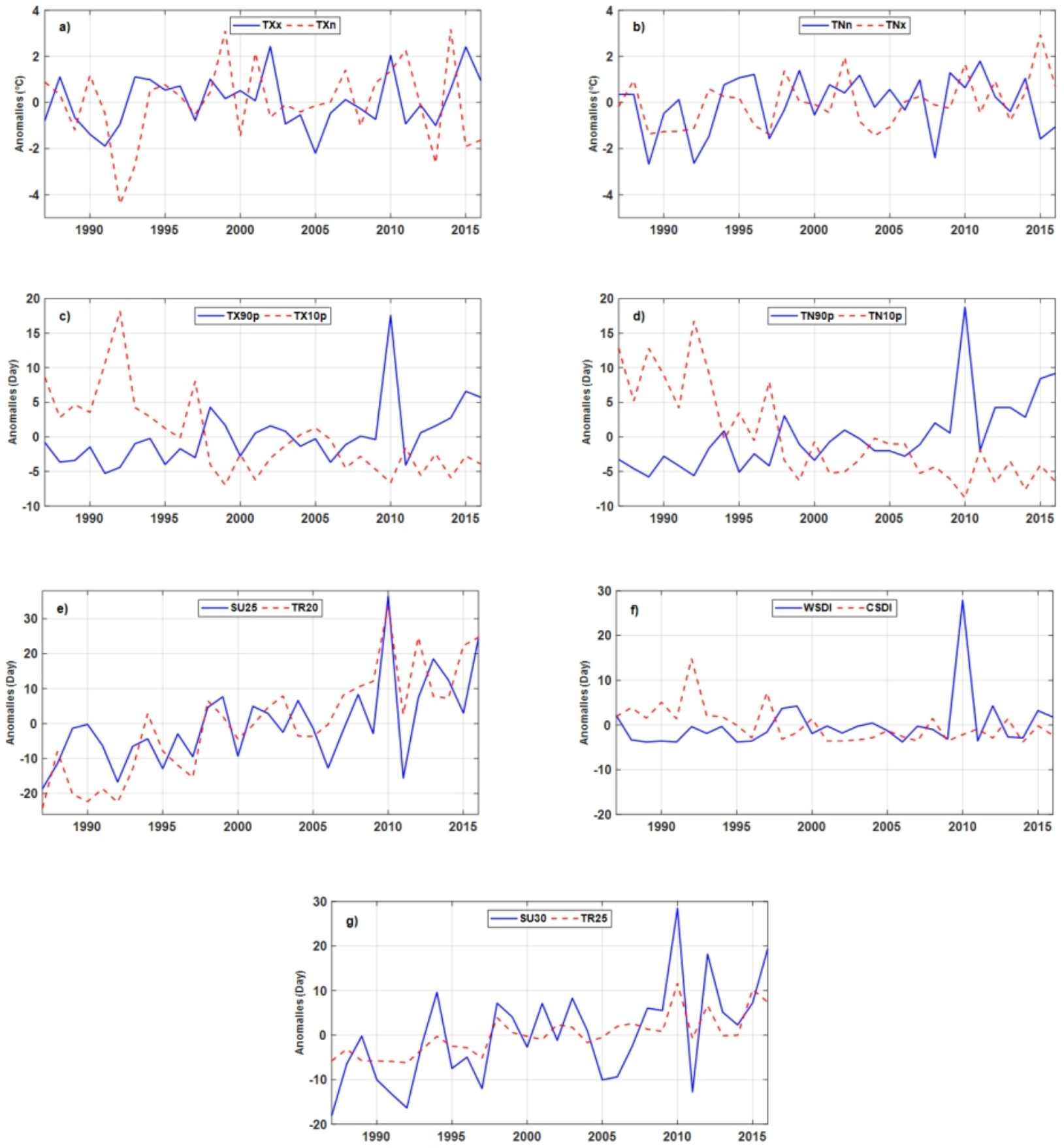

Figure 2. Time series of anomalies for the annual extreme temperature indices averaged over the study area during the period 1987–2016.

437x412mm (96 x 96 DPI)

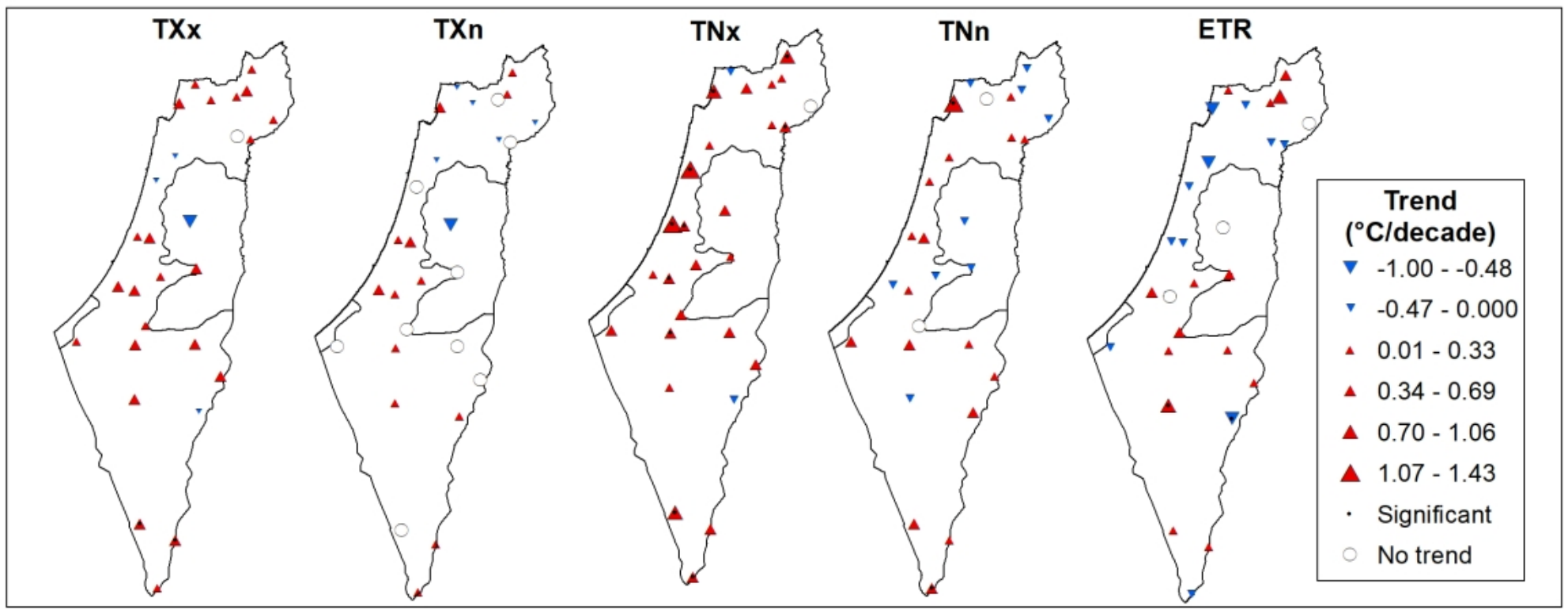

Figure 3. Spatial distribution of trends (ºC/decade) for absolute extreme temperature indices.

637x254mm (96 x 96 DPI)

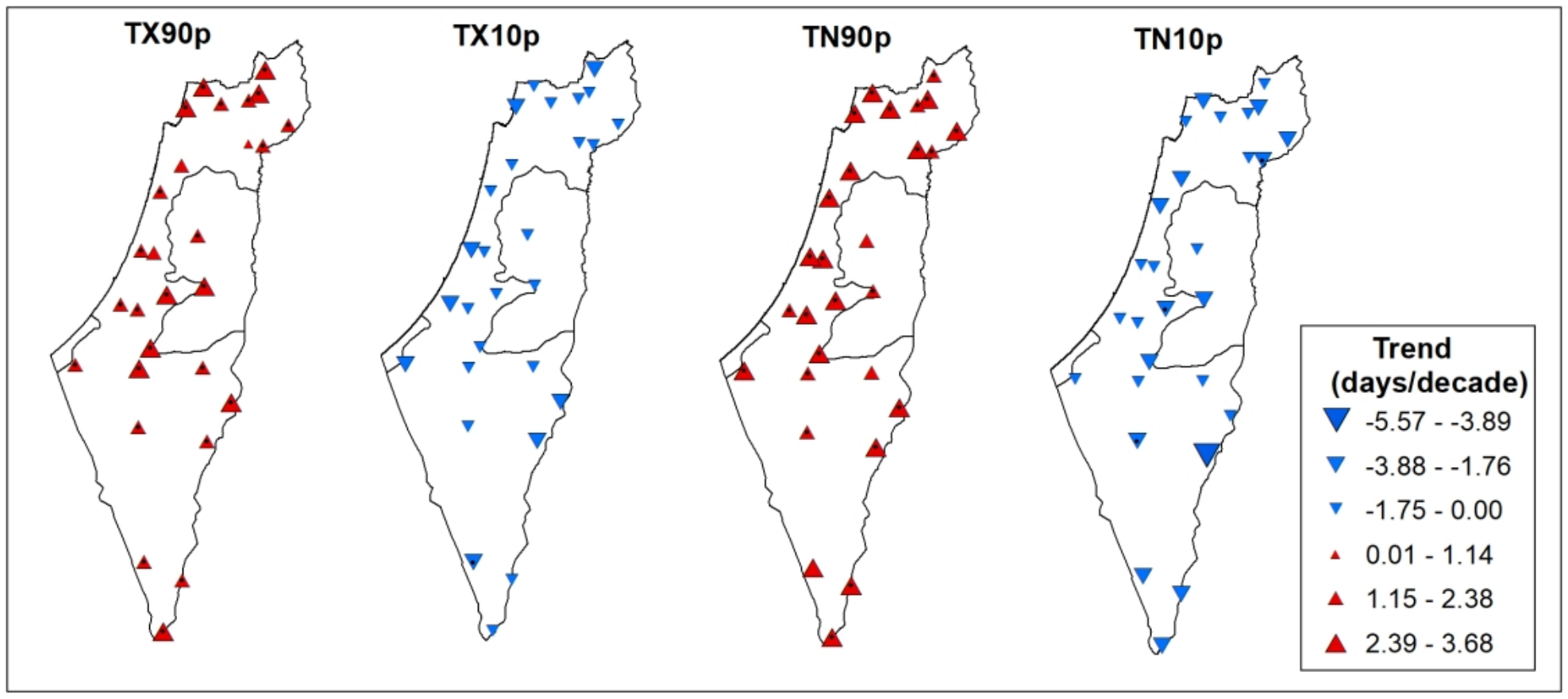

Figure 4. Spatial distribution of trends (days/decade) for annual percentile-based extreme temperature indices.

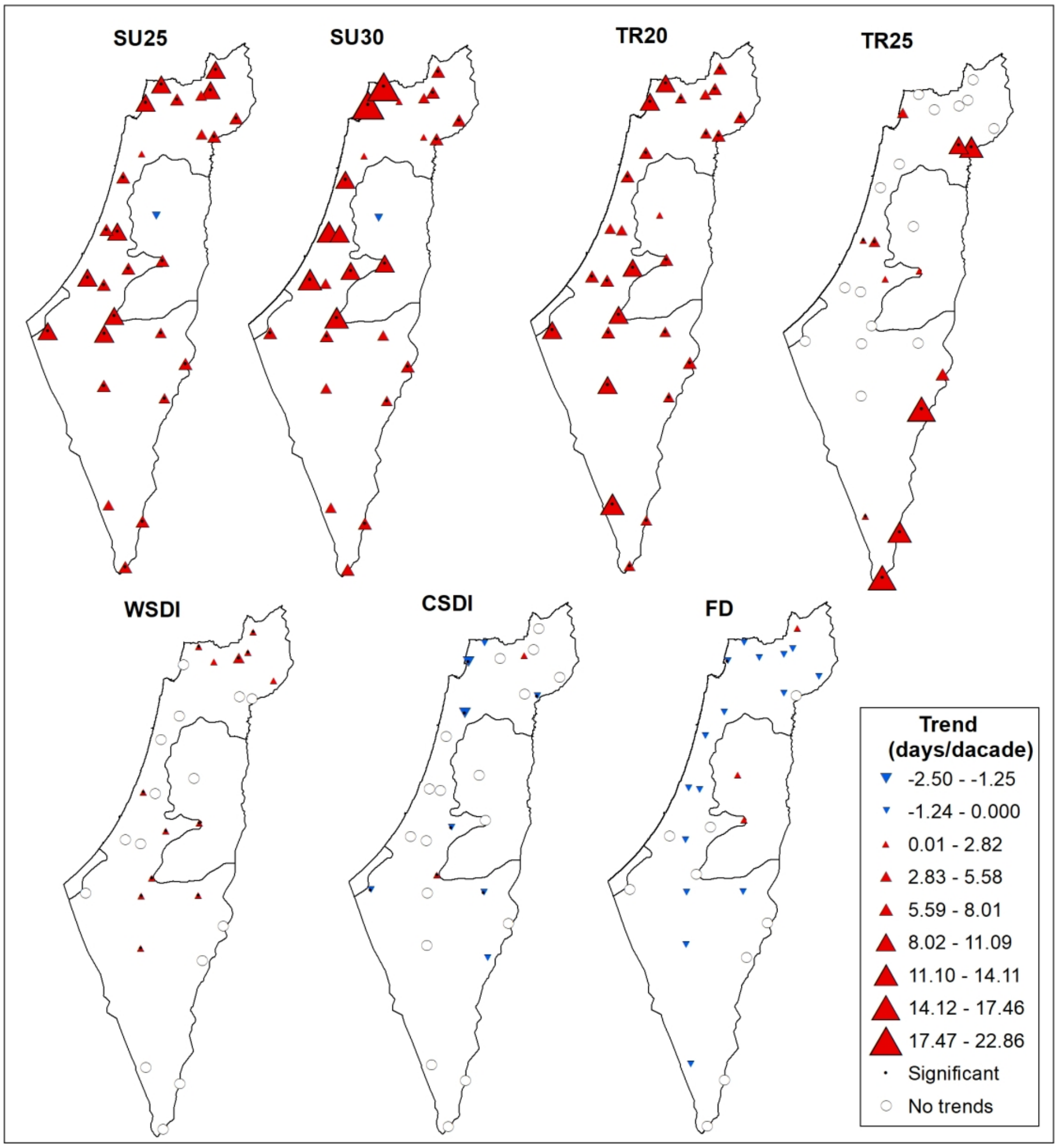

Figure 5. Spatial distribution of trends (days/decade) for annual duration and fixed threshold extreme temperature indices.

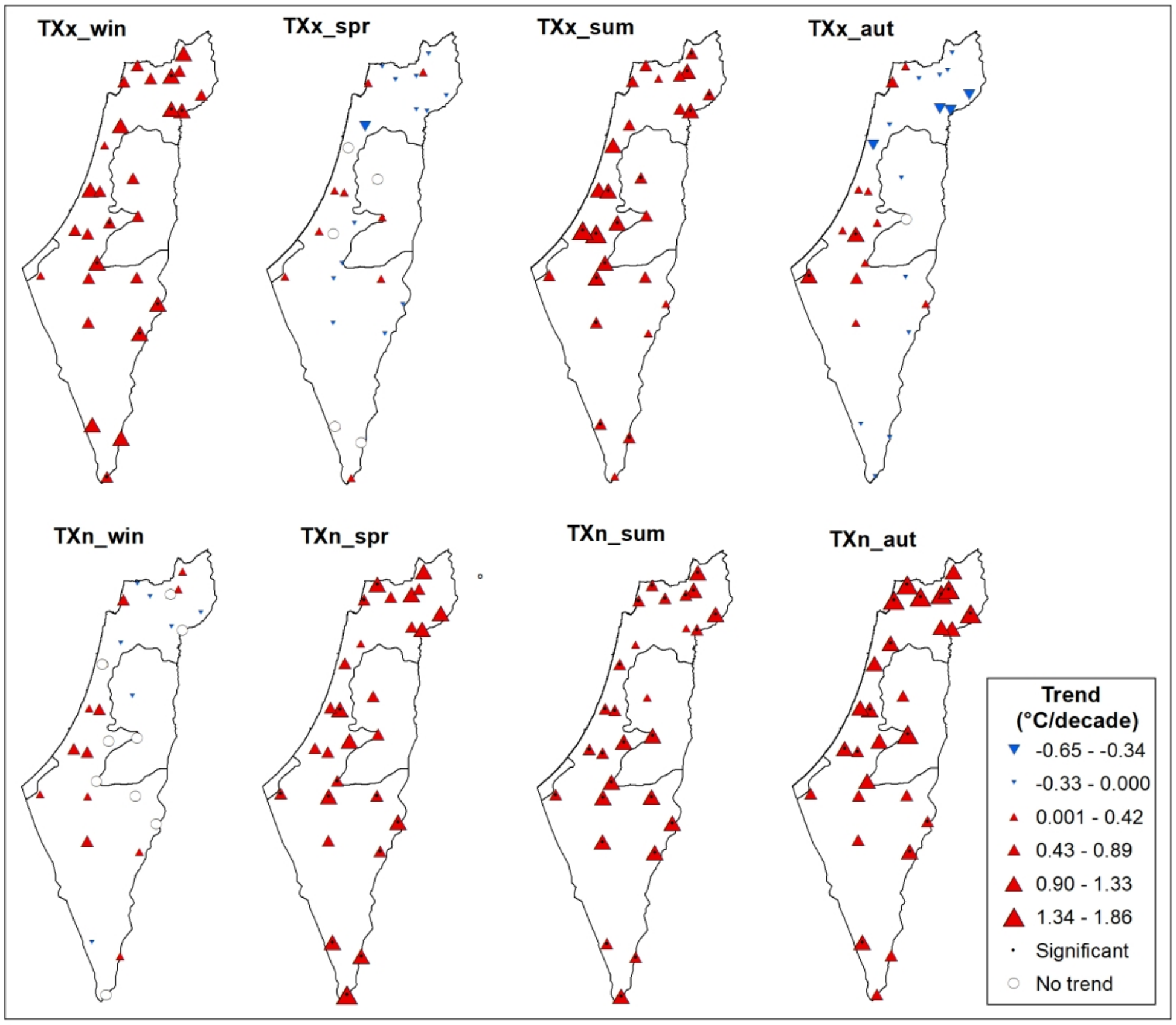

Figure 6. Spatial distribution of trends (ºC/decade) for seasonal absolute extreme maximum temperature indices.

599x519mm (96 x 96 DPI)

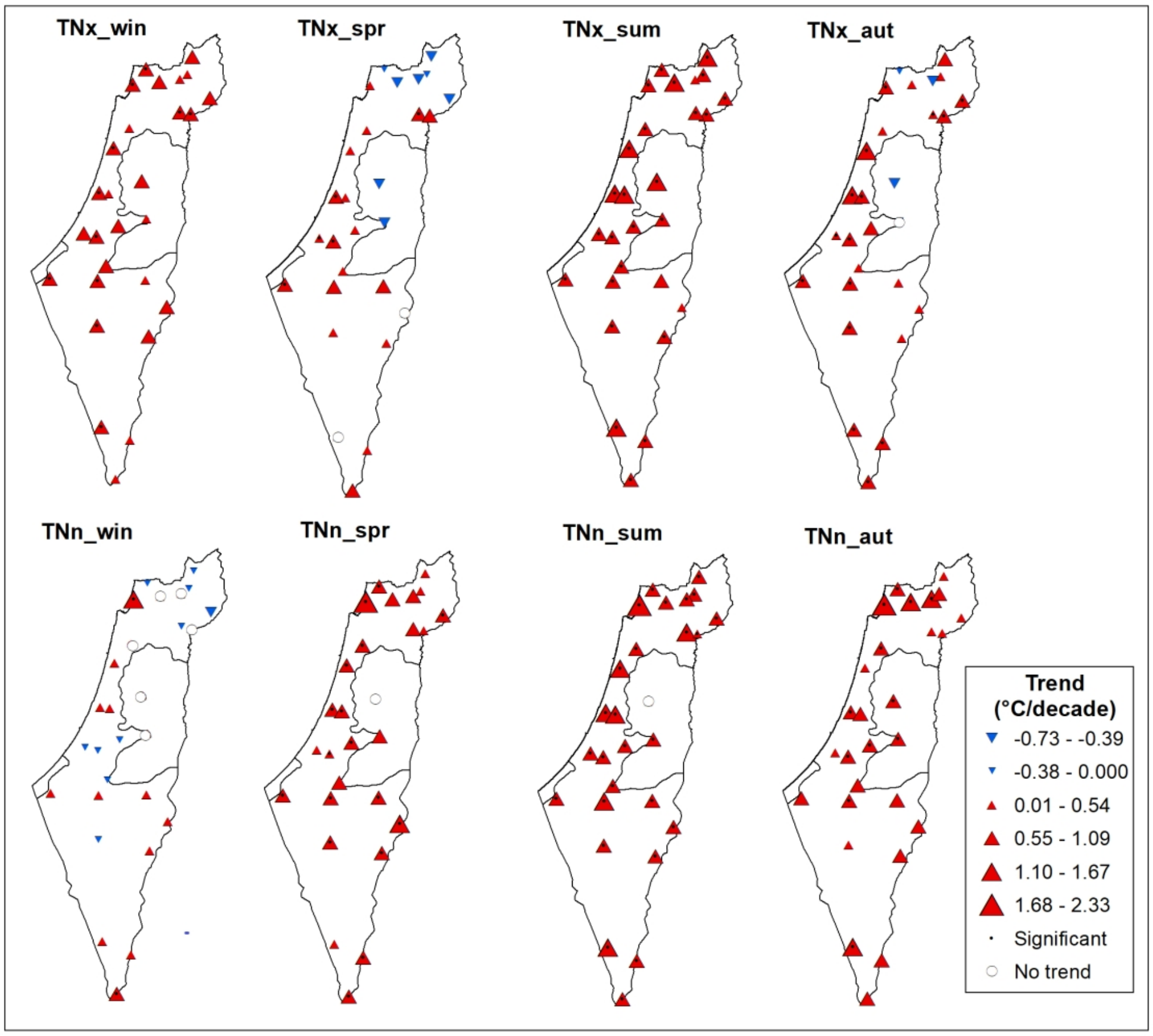

Figure 7. Spatial distribution of trends (oC/decade) for seasonal absolute extreme minimum temperature indices.

590x524mm (96 x 96 DPI)

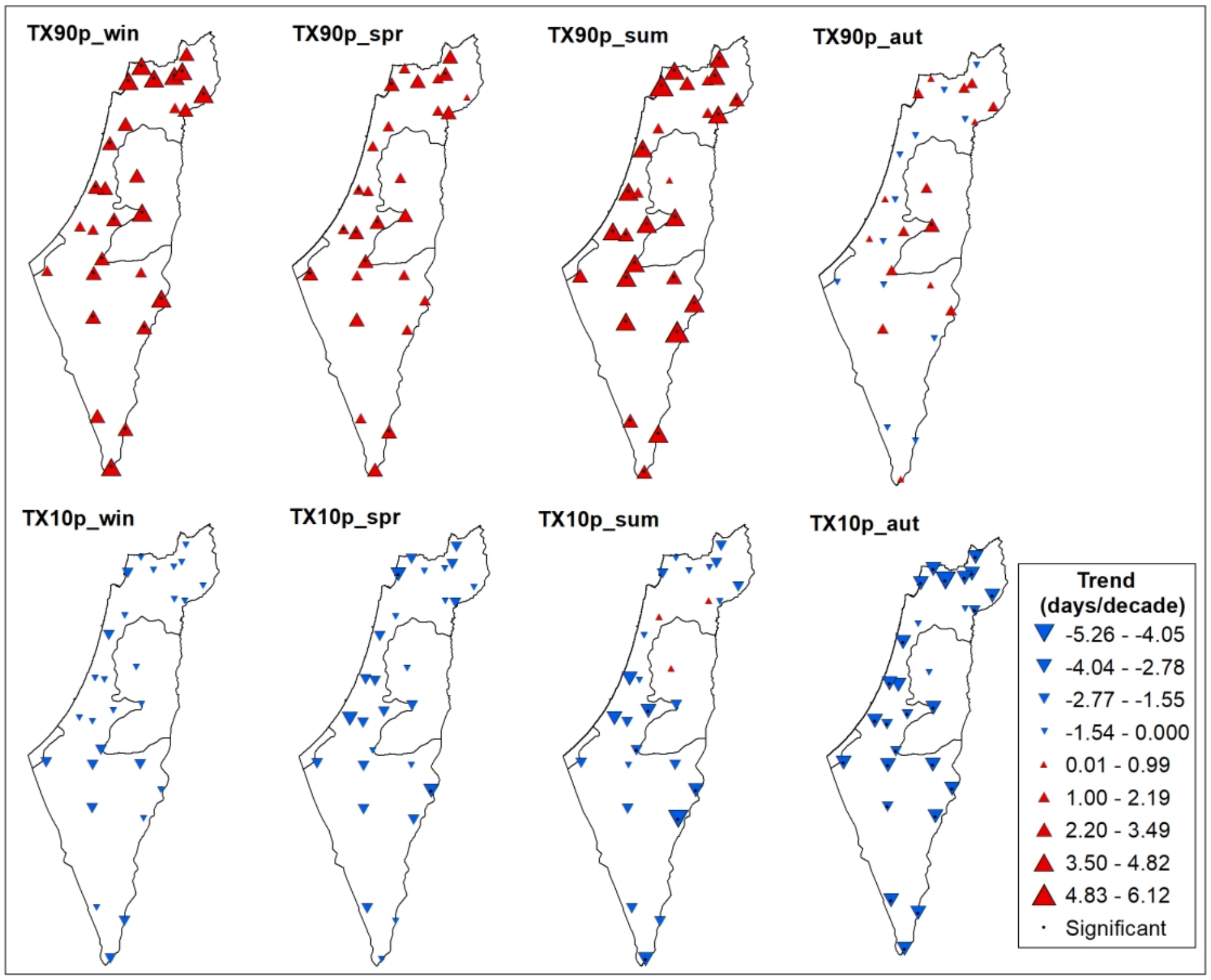

Figure 8. Spatial distribution of trends (days/decade) for seasonal percentile-based extreme temperature indices.

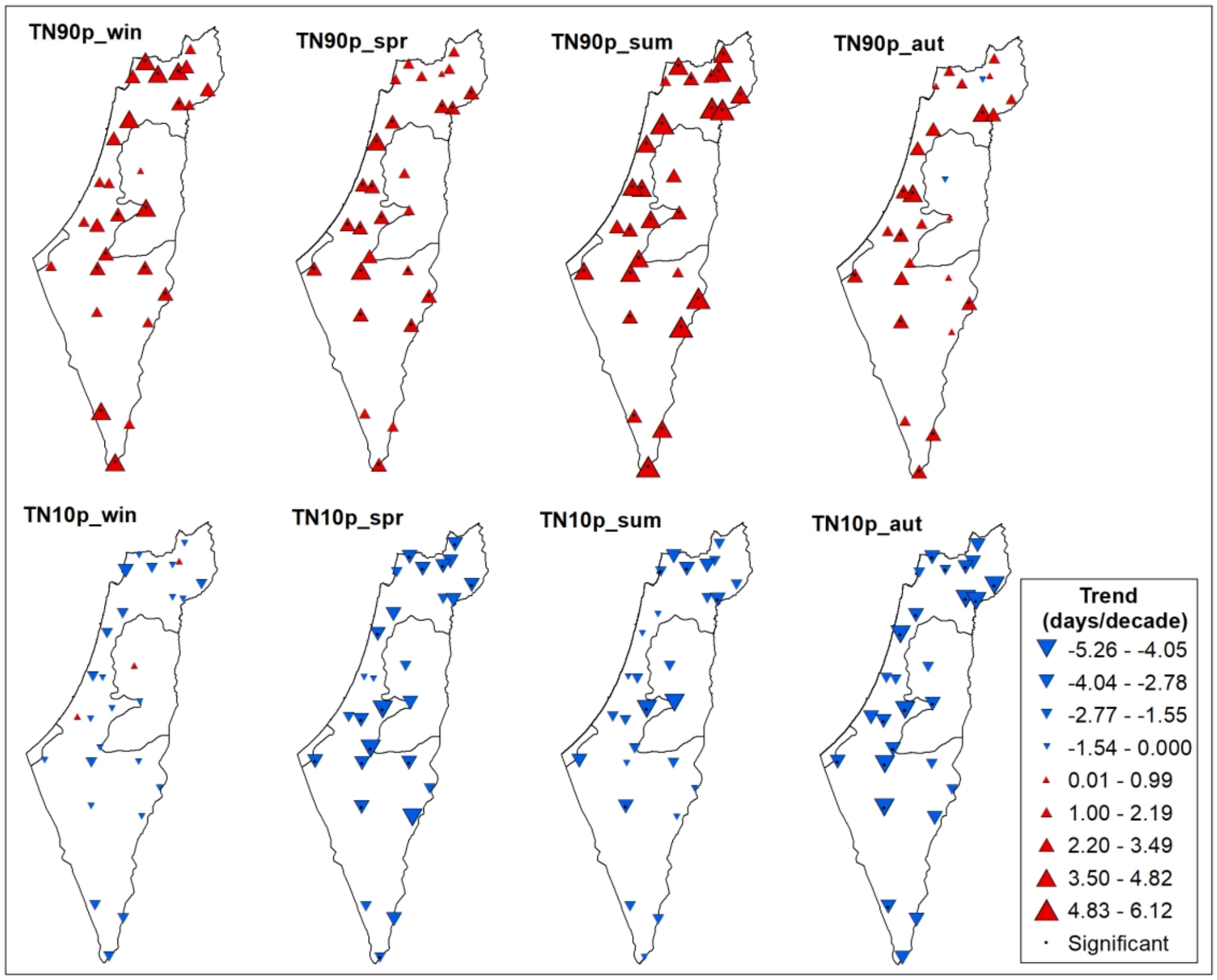

Figure 9. Spatial distribution of trends (days/decade) for seasonal percentile-based extreme temperature indices.

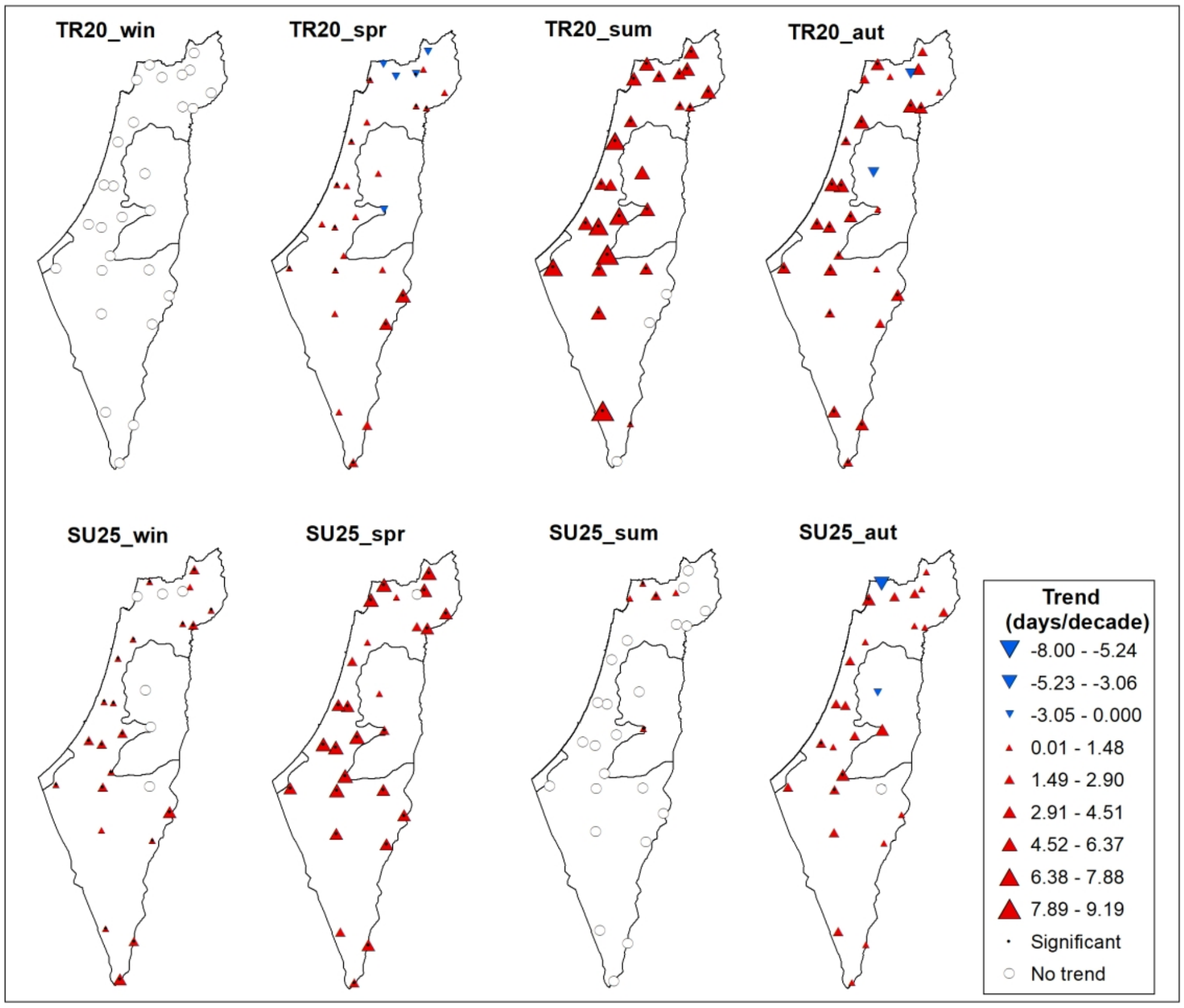

Figure 10. Spatial distribution of trends (days/decade) for duration and fixed threshold extreme temperature indices.

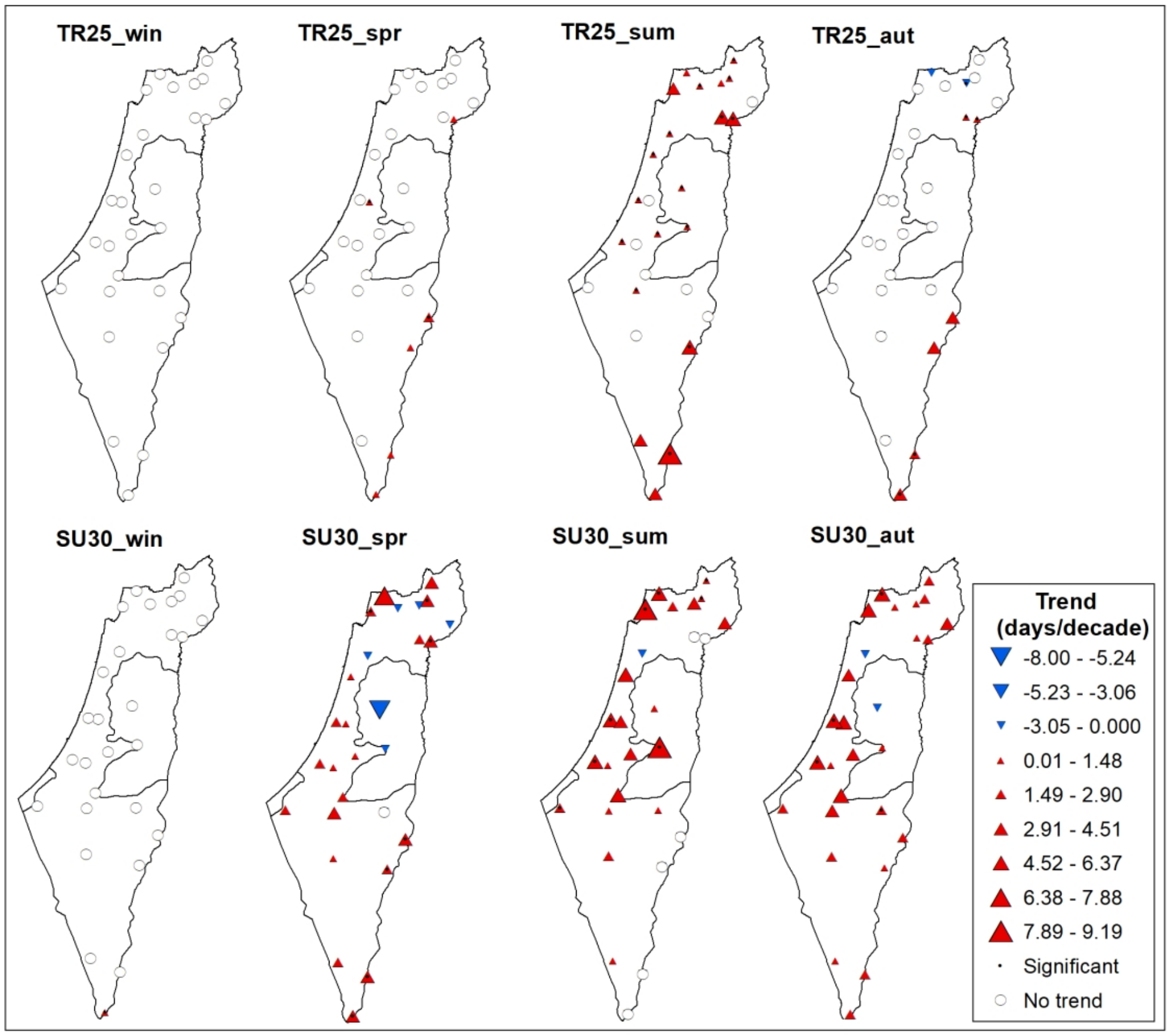

Figure 11. Spatial distribution of trends (days/decade) for duration and fixed threshold extreme temperature indices.

576x513mm (96 x 96 DPI)

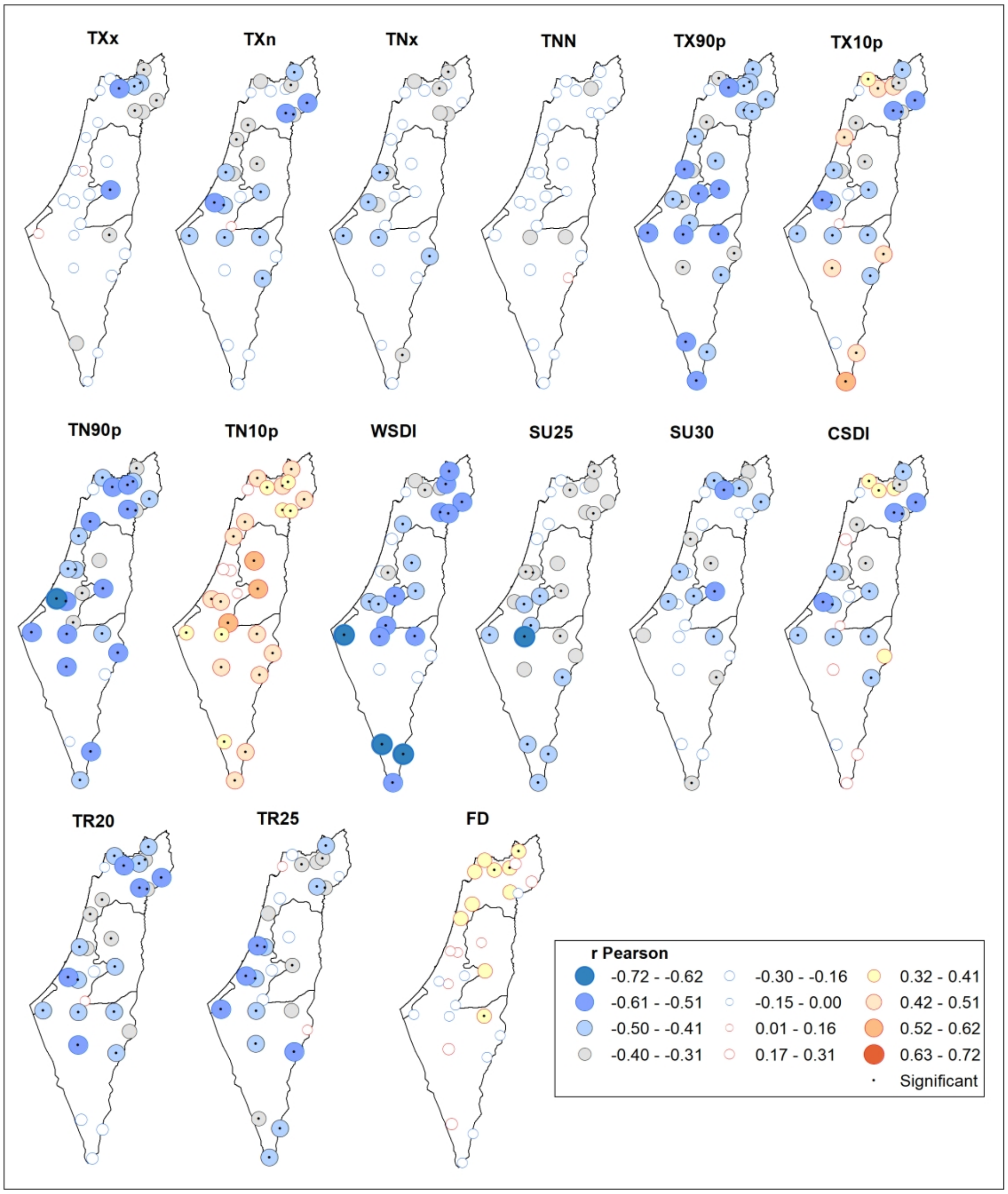

Figure 12. Spatial distribution of Pearson correlation coefficients between the NCP index and the extreme temperature indices at annual scale.

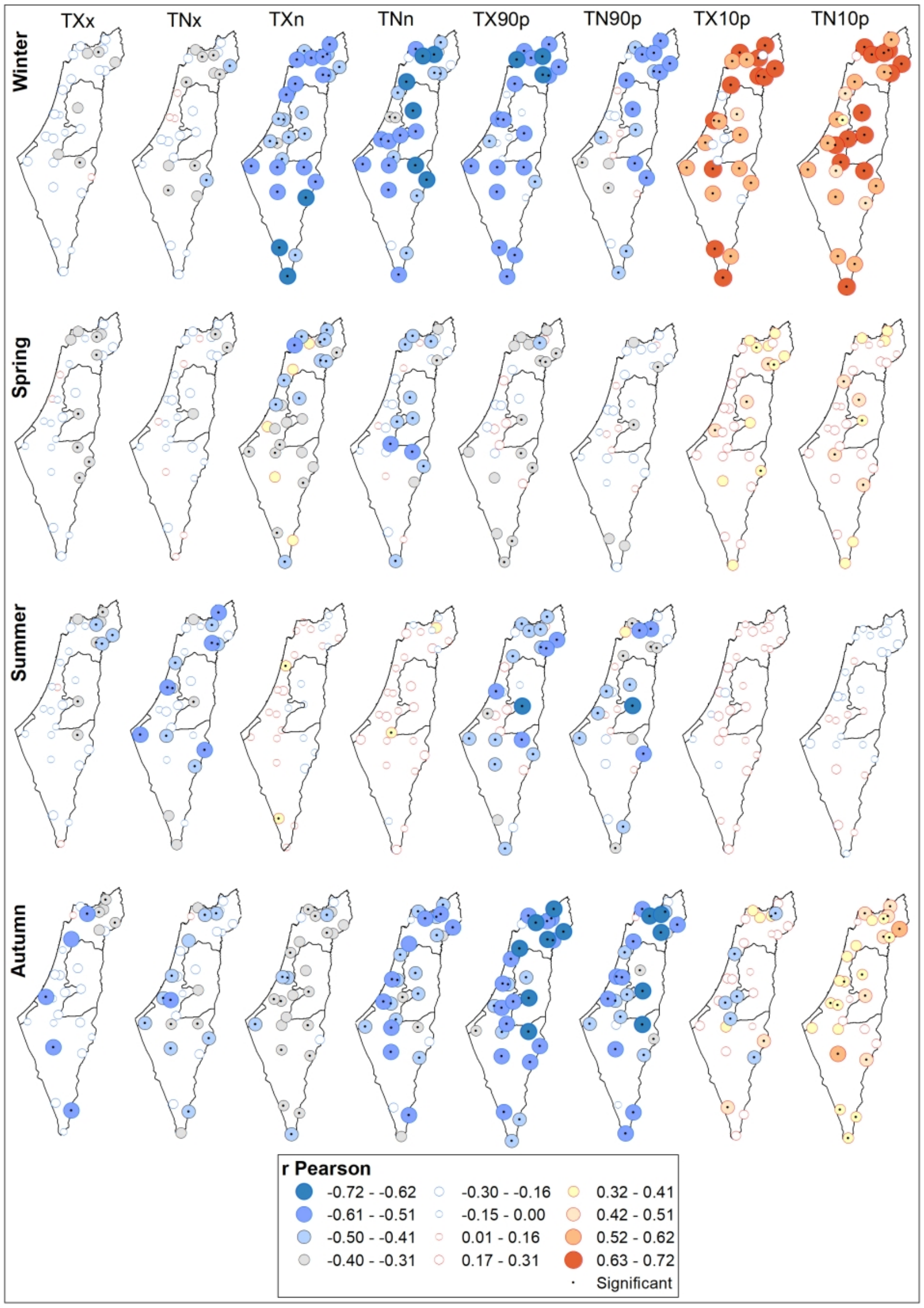

Figure 13. Spatial distribution of Pearson correlation coefficients between the NCP index and the extreme temperature indices at seasonal scale.

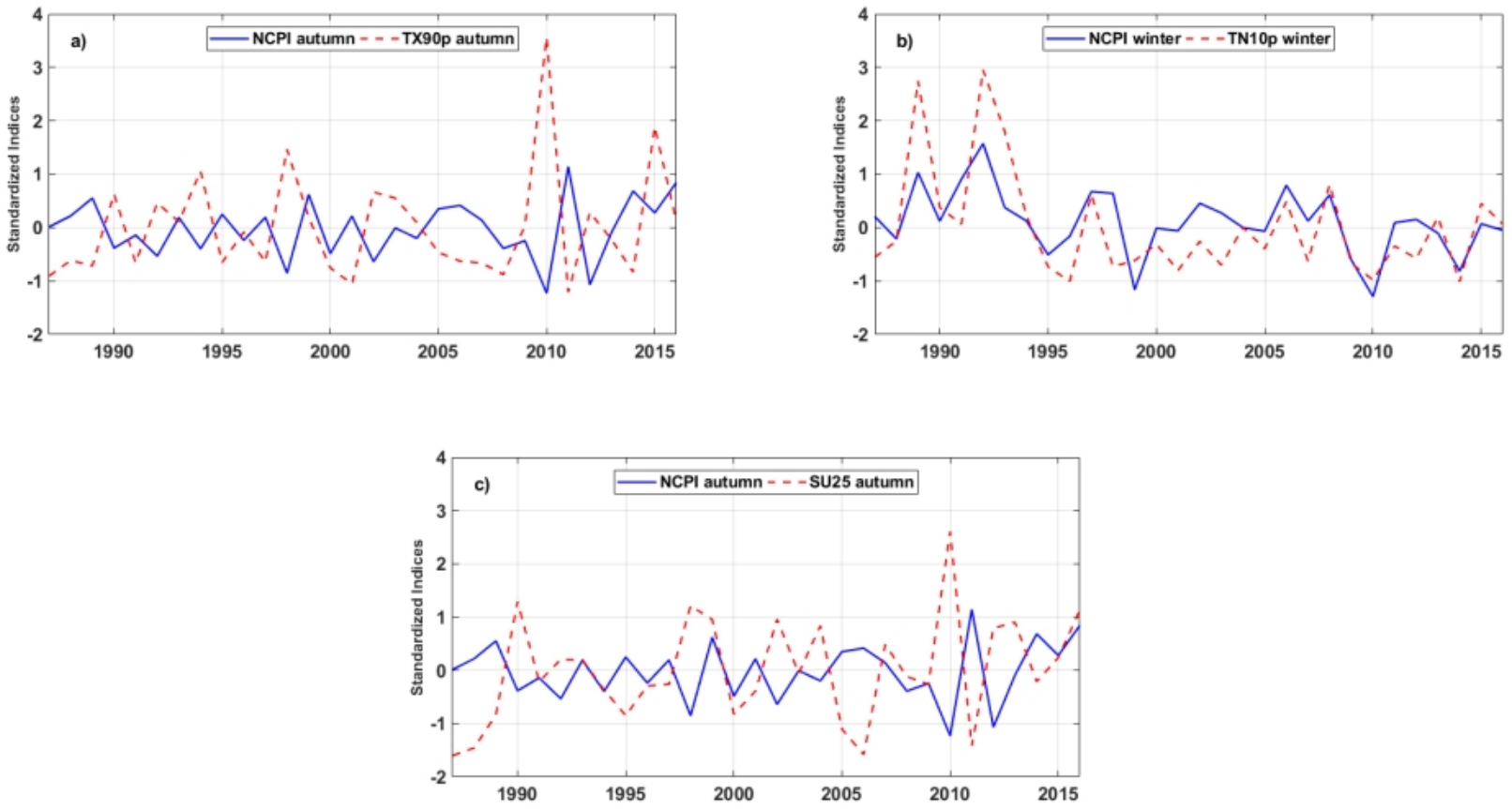

Figure 14. Temporal variability of the standardized atmospheric/oceanic mode index with some standardized temperature indices time series.

Table 1. Meteorological stations used in this study.

| Station | Name | Elev. (m) | Longitude (E) | Latitude (N) | Data Period | Missing values % | Total Outliers |
|---|---|---|---|---|---|---|---|
| **1** | Elat | 12 | 29.5526$^0$ | 34.9542$^0$ | 1987-2016 | 0 | 0 |
| **2** | Yotvata | 70 | 29.8851$^0$ | 35.0771$^0$ | 1987-2016 | 1.3 | 0 |
| **3** | Sede Boqer | 475 | 30.8704$^0$ | 34.7951$^0$ | 1987-2016 | 0 | 0 |
| **4** | Sedom Man | -388 | 31.0306$^0$ | 35.3919$^0$ | 1987-2016 | 0 | 0 |
| **5** | Arad Man | 605 | 31.2552$^0$ | 35.2128$^0$ | 1987-2016 | 0 | 4 |
| **6** | Beer Sheva Man | 279 | 31.2515$^0$ | 34.7995$^0$ | 1987-2016 | 0 | 0 |
| **7** | Besor Farm Man | 110 | 31.2716$^0$ | 34.3894$^0$ | 1987-2016 | 0.3 | 7 |
| **8** | Lahav Man | 460 | 31.3812$^0$ | 34.8729$^0$ | 1987-2016 | 0 | 2 |
| **9** | Lod Airport | 40 | 31.9950$^0$ | 34.8970$^0$ | 1987-2016 | 0.3 | 5 |
| **10** | Gat Man | 140 | 31.6303$^0$ | 34.7913$^0$ | 1987-2016 | 0.1 | 5 |
| **11** | Negba | 90 | 31.6596$^0$ | 34.6796$^0$ | 1987-2016 | 0 | 8 |
| **12** | Beit Jimal Man | 355 | 31.7248$^0$ | 34.9762$^0$ | 1987-2016 | 4.1 | 4 |
| **13** | Jerusalem Center | 810 | 31.7806$^0$ | 35.2217$^0$ | 1987-2016 | 0 | 0 |
| **14** | Bet Dagan Man | 31 | 32.0073$^0$ | 34.8138$^0$ | 1987-2016 | 0 | 10 |
| **15** | Yavne’el Man | 0 | 32.6978$^0$ | 35.5101$^0$ | 1987-2016 | 0 | 5 |
| **16** | Hakfar Hayarok | 432 | 30.0000$^0$ | 34.8300$^0$ | 1987-2016 | 0.3 | 8 |
| **17** | Massada Man | -200 | 32.6833$^0$ | 35.6008$^0$ | 1987-2016 | 4.1 | 0 |
| **18** | Avne Etan Man | 375 | 32.8174$^0$ | 35.7622$^0$ | 1987-2016 | 0.5 | 0 |
| **19** | Kefar Blum Man | 75 | 33.1728$^0$ | 35.6120$^0$ | 1987-2016 | 0 | 1 |
| **20** | Ayyelet Hashahar | 175 | 33.0219$^0$ | 35.5742$^0$ | 1987-2016 | 0 | 2 |
| **21** | Zefat Har Kennan | 936 | 32.9800$^0$ | 35.5070$^0$ | 1987-2016 | 0.7 | 0 |
| **22** | Harashim | 825 | 32.9560$^0$ | 35.3288$^0$ | 1987-2016 | 0.1 | 5 |
| **23** | Elon Man | 300 | 33.0653$^0$ | 35.2173$^0$ | 1987-2016 | 0 | 9 |
| **24** | En Hahoresh Man | 15 | 32.3877$^0$ | 34.9376$^0$ | 1987-2016 | 2 | 16 |
| **25** | Galed (Even Yizhaq) | 185 | 32.5580$^0$ | 35.0742$^0$ | 1987-2016 | 0 | 8 |
| **26** | Akko | 8 | 32.9318$^0$ | 35.1020$^0$ | 1987-2016 | 0 | 7 |
| **27** | Hazeva Man | -135 | 30.7787$^0$ | 35.2389$^0$ | 1988-2016 | 0.8 | 3 |
| **28** | Ariel | 590 | 32.1063$^0$ | 35.1774$^0$ | 1990-2016 | 0.5 | 9 |

Table 2. Description of extreme temperature indices used in this study.

| Index | Indicator name | Description | Index type | Unit |
|---|---|---|---|---|
| **TXx** | Max Tmax | Maximum value of daily maximum temperature | Absolut | $^0$C |
| **TXn** | Min Tmax | Minimum value of daily maximum temperature | Absolut | $^0$C |
| **TNx** | Max Tmin | Maximum value of daily minimum temperature | Absolut | $^0$C |
| **TNn** | Min Tmin | Minimum value of daily minimum temperature | Absolut | $^0$C |
| **ETR** | Range | MaxTmax – MinTmin | Absolut | $^0$C |
| **TX90p** | Warm days | Percentage of days with Tmax > 90th percentile | Percentile | Days |
| **TX10p** | Cold days | Percentage of days with Tmax < 10th percentile | Percentile | Days |
| **TN90p** | Warm nights | Percentage of days with Tmin > 90th percentile | Percentile | Days |
| **TN10p** | Cold nights | Percentage of days with Tmin < 10th percentile | Percentile | Days |
| **WSDI** | Warm spell duration index | Number of days with at least six consecutive days with Tmax > 90th percentile | Percentile Duration | Days |
| **CSDI** | Cold spell duration index | Number of days with at least six consecutive days with Tmin < 10th percentile | Percentile Duration | Days |
| **SU25** | Summer days | Number of days with Tmax > 25 $^0$C | Threshold | Days |
| **SU30** | Very summer days | Number of days with Tmax > 30 $^0$C | Threshold | Days |
| **FD** | Frost days | Number of days with Tmin < 0$^0$C | Threshold | Days |
| **TR20** | Tropical nights | Number of days with Tmin > 20$^0$C | Threshold | Days |
| **TR25** | Very tropical nights | Number of days with Tmin > 25$^0$C | Threshold | Days |

Table 3. Trends for the annual extreme temperature indices averaged over the study area. Symbols ***, **, * and +, indicate a significant trend at α = 0.001, α = 0.01, α = 0.05 and α = 0.1 level, respectively.

| Index | Trend | Index | Trend |
|---|---|---|---|
| TXx | 0.17 ($^0$C/decade) | TN10p | -5.52 (days/decade) |
| TXn | 0.05 ($^0$C/decade) | WSDI | 0.63* (days/decade) |
| TNx | 0.40+ ($^0$C/decade) | SU25 | 8.15*** (days/decade) |
| TNn | 0.18 ($^0$C/decade) | SU30 | 7.10* (days/decade) |
| ETR | -0.01 ($^0$C/decade) | FD | 0 (days/decade) |
| TX90p | 2.20** (days/decade) | CSDI | -1.79** (days/decade) |
| TX10p | -3.94 (days/decade) | TR20 | 14.40*** (days/decade) |
| TN90p | 3.17*** (days/decade) | TR25 | 3.74** (days/decade) |

Table 4. Number of stations showing positive and negative trends for each extreme temperature index at annual scale. In bracket the number of stations showing significant trends at the 95% confidence level. Right column shows the percentage of stations with significant increasing/decreasing trends.

| Index | Positive | Negative | Percentage of stations (%) |
|---|---|---|---|
| TXx | 23 (2) | 4 (0) | (7.1/ 0) |
| TXn | 13 (0) | 6 (0) | (0 / 0) |
| TNn | 17 (2) | 9 (0) | (7.1/ 0) |
| TNx | 25(10) | 2 (0) | (36/ 0) |
| ETR | 14 (1) | 11 (1) | (3.5/ 3.5) |
| TX90p | 28 (25) | 0 | (89/0) |
| TX10p | 0 | 28 (1) | (0/3.5) |
| TN90p | 28 (26) | 0 | (93/0) |
| TN10p | 0 | 28 (3) | (0/10.7) |
| WSDI | 13 (11) | 0 | (39/0) |
| CSDI | 2 (1) | 8(6) | (0/21) |
| SU25 | 27 (23) | 1 (0) | (82/0) |
| SU30 | 26 (16) | 1 (0) | (57/0) |
| FD | 3 (0) | 16 (1) | (0/5) |
| TR20 | 28 (24) | 0 | (86/0) |
| TR25 | 12 (8) | 0 | (28.5/0) |

Table 5. Number of stations with positive and negative trends for each extreme temperature index at seasonal scale. In bracket the number of stations with significant positive or negative trends at the 95% confidence level.

| Index | Seasonal scale | | | | | | | |
|---|---|---|---|---|---|---|---|---|
| | Winter | | Spring | | Summer | | Autumn | |
| | Positive (+) | Negative (-) | Positive (+) | Negative (-) | Positive (+) | Negative (-) | Positive (+) | Negative (-) |
| TXx | 28 (8) | 0 | 9(0) | 14 (0) | 28(14) | 0 | 12(2) | 15(0) |
| TXn | 12(0) | 7(0) | 28(12) | 0 | 28(25) | 0 | 28(14) | 0 |
| TNn | 12(2) | 10(0) | 27(17) | 0 | 27(27) | 0 | 28(12) | 0 |
| TNx | 28(11) | 0 | 18(5) | 8(0) | 28(25) | 0 | 24(15) | 3(0) |
| TX90p | 28(17) | 0 | 28(9) | 0 | 28(21) | 0 | 16(1) | 12(0) |
| TX10p | 0 | 28(0) | 0 | 28(2) | 3(0) | 25(5) | 0 | 28(24) |
| TN90p | 28(12) | 0 | 28(17) | 0 | 28(24) | 0 | 26(9) | 2(0) |
| TN10p | 3(0) | 25(0) | 0 | 28(14) | 0 | 28(5) | 0 | 28(16) |
| SU25 | 22 (20) | 0 | 27 (21) | 0 | 5 (3) | 0 | 25 (4) | 2 (0) |
| SU30 | 1 (1) | 0 | 21 (7) | 6 (0) | 21 (8) | 1 (0) | 26 (4) | 2 (0) |
| TR20 | 0 | 0 | 23 (11) | 5 (1) | 25 (23) | 0 | 26 (18) | 2 (0) |
| TR25 | 0 | 0 | 6 (2) | 0 | 20 (15) | 0 | 6 (4) | 2 (1) |

Table 6. Number of stations with significant positive or negative correlations between extreme temperature and teleconnection indices at annual scale. Only significant results at the 95% confidence level are shown.

| Index | NAO | | EA | | EA/WR | | WEMO | | MOI | | NCPI | | ENSO | |
|---|---|---|---|---|---|---|---|---|---|---|---|---|---|---|
| | (+) | (-) | (+) | (-) | (+) | (-) | (+) | (-) | (+) | (-) | (+) | (-) | (+) | (-) |
| **TXx** | 1 | 0 | 8 | 0 | 0 | 2 | 5 | 0 | 0 | 2 | 0 | 8 | 0 | 0 |
| **TXn** | 0 | 1 | 1 | 0 | 0 | 0 | 0 | 3 | 0 | 0 | 0 | 16 | 0 | 0 |
| **TNx** | 10 | 1 | 11 | 0 | 1 | 0 | 0 | 0 | 0 | 0 | 0 | 9 | 0 | 0 |
| **TNn** | 0 | 1 | 1 | 1 | 0 | 1 | 0 | 1 | 0 | 0 | 0 | 0 | 0 | 3 |
| **TX90p** | 0 | 17 | 8 | 0 | 0 | 12 | 0 | 0 | 0 | 5 | 0 | 26 | 0 | 0 |
| **TX10p** | 1 | 0 | 0 | 0 | 12 | 0 | 0 | 0 | 1 | 0 | 8 | 15 | 4 | 0 |
| **TN90p** | 0 | 14 | 7 | 0 | 0 | 14 | 0 | 0 | 0 | 4 | 0 | 24 | 0 | 0 |
| **TN10p** | 2 | 0 | 0 | 0 | 11 | 0 | 0 | 9 | 0 | 0 | 21 | 0 | 14 | 0 |
| **WSDI** | 0 | 0 | 1 | 0 | 0 | 12 | 1 | 1 | 9 | 0 | 0 | 20 | 0 | 0 |
| **SU25** | 0 | 7 | 1 | 0 | 0 | 24 | 3 | 0 | 0 | 0 | 0 | 16 | 0 | 0 |
| **SU30** | 0 | 7 | 6 | 0 | 0 | 12 | 5 | 0 | 0 | 0 | 0 | 13 | 0 | 0 |
| **CSDI** | 0 | 1 | 0 | 1 | 12 | 1 | 3 | 0 | 0 | 0 | 0 | 18 | 3 | 0 |
| **TR20** | 0 | 8 | 7 | 0 | 0 | 9 | 4 | 0 | 0 | 0 | 0 | 20 | 0 | 2 |
| **TR25** | 0 | 8 | 6 | 0 | 0 | 4 | 0 | 0 | 1 | 0 | 0 | 17 | 0 | 0 |
| **FD** | 0 | 0 | 0 | 0 | 5 | 0 | 5 | 1 | 0 | 0 | 4 | 0 | 0 | 0 |

Table 7. Number of stations with significant correlations between extreme temperature and teleconnection indices at seasonal scale. Only significant results at the 95% confidence level are shown.

| Indices | Seasons | Extreme temperature indices | | | | | | | | | | |
|---|---|---|---|---|---|---|---|---|---|---|---|---|
| | | TXx | TNx | TXn | TNn | TX 90p | TN 90p | TX 10p | TN 10p | SU 25/30 | TR 20/25 | Tot. |
| NAO | Winter | 3 | 11 | 2 | 6 | 1 | 0 | 6 | 9 | 20/19 | 0/0 | 77 |
| | Spring | 0 | 8 | 0 | 1 | 4 | 2 | 0 | 4 | 19/7 | 5/0 | 50 |
| | Summer | 0 | 0 | 0 | 1 | 13 | 11 | 1 | 1 | 2/3 | 4/3 | 39 |
| | Autumn | 0 | 5 | 25 | 5 | 6 | 2 | 0 | 0 | 1/2 | 6/2 | 54 |
| EA | Winter | 0 | 0 | 0 | 4 | 0 | 0 | 0 | 0 | 0/0 | 0/0 | 4 |
| | Spring | 1 | 6 | 0 | 1 | 4 | 6 | 0 | 2 | 1/5 | 11/6 | 43 |
| | Summer | 6 | 5 | 0 | 7 | 0 | 5 | 1 | 0 | 0/2 | 0/3 | 29 |
| | Autumn | 0 | 1 | 0 | 1 | 0 | 0 | 0 | 0 | 14/4 | 0/0 | 20 |
| EA/WR | Winter | 0 | 12 | 19 | 12 | 15 | 2 | 17 | 16 | 11/0 | 0/0 | 104 |
| | Spring | 27 | 6 | 23 | 3 | 6 | 10 | 2 | 7 | 22/14 | 14/8 | 142 |
| | Summer | 16 | 17 | 28 | 25 | 1 | 9 | 3 | 0 | 2/2 | 5/6 | 114 |
| | Autumn | 27 | 21 | 18 | 0 | 0 | 0 | 0 | 2 | 0/2 | 0/0 | 70 |
| WEMO | Winter | 0 | 0 | 7 | 18 | 0 | 1 | 0 | 0 | 0/0 | 0/0 | 26 |
| | Spring | 2 | 10 | 0 | 0 | 10 | 7 | 16 | 4 | 24/12 | 3/0 | 88 |
| | Summer | 0 | 0 | 1 | 2 | 0 | 0 | 1 | 1 | 2/2 | 1/1 | 11 |
| | Autumn | 3 | 0 | 15 | 20 | 8 | 3 | 4 | 0 | 20/2 | 1/1 | 74 |
| MOI | Winter | 0 | 6 | 5 | 5 | 22 | 16 | 1 | 3 | 17/8 | 0/0 | 83 |
| | Spring | 1 | 0 | 1 | 2 | 0 | 0 | 1 | 0 | 14/4 | 4/0 | 27 |
| | Summer | 1 | 0 | 1 | 3 | 0 | 0 | 5 | 1 | 2/5 | 2/0 | 20 |
| | Autumn | 9 | 2 | 7 | 0 | 0 | 0 | 0 | 0 | 6/0 | 0/10 | 34 |
| NCPI | Winter | 2 | 8 | 28 | 24 | 21 | 17 | 22 | 27 | 4/3 | 0/0 | 156 |
| | Spring | 6 | 2 | 17 | 11 | 9 | 2 | 6 | 5 | 19/14 | 2/0 | 93 |
| | Summer | 7 | 12 | 2 | 1 | 17 | 14 | 0 | 0 | 0/4 | 5/9 | 71 |
| | Autumn | 7 | 11 | 17 | 22 | 27 | 22 | 9 | 17 | 28/19 | 21/4 | 204 |
| ENSO | Winter | 1 | 0 | 2 | 3 | 0 | 1 | 9 | 1 | 2/2 | 1/0 | 22 |
| | Spring | 0 | 0 | 0 | 0 | 0 | 0 | 19 | 17 | 4/1 | 2/0 | 41 |
| | Summer | 0 | 0 | 2 | 8 | 0 | 2 | 17 | 15 | 5/10 | 10/3 | 72 |
| | Autumn | 0 | 0 | 0 | 7 | 1 | 1 | 1 | 0 | 0/0 | 0/5 | 15 |

**Spatio-temporal Analysis for Extreme Temperature Indices over Levant region**

Ala A. M. Salameh, Sonia Raquel Gámiz-Fortis, Yolanda Castro-Díez, Ahmad Abu Hammad and María Jesús Esteban-Parra*

Trends of 16 climate extreme indices based on daily maximum and minimum temperatures during the period 1987–2016 at 28 stations across the Levant region were evaluated.

A dominant warming for the last three decades, with more intense changes for minimum than for maximum temperatures. Significant increasing trends for summer days (SU25) and tropical nights (TR20) were found.

North Sea-Caspian pattern (NCP) is the main driver of extreme temperature indices over the study area.

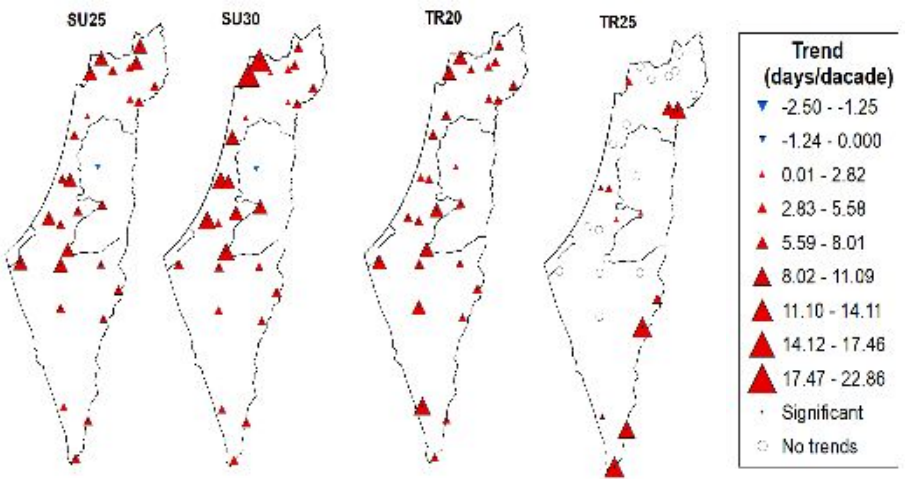